\RequirePackage{fix-cm}  
\documentclass[prapplied,twocolumn,floatfix,superscriptaddress,longbibliography]{revtex4-2}
\usepackage{amssymb,amsmath,amstext}
\usepackage{graphicx}
\usepackage{subfigure}
\usepackage{amsthm}
\usepackage{verbatim}
\usepackage{dcolumn}
\usepackage{bm}
\usepackage{epsf}
\usepackage{xcolor}
\usepackage[colorlinks=true,citecolor=blue,linkcolor=blue,urlcolor=blue]{hyperref}
\newcommand{\conf}[2]{{\underline{#1}}_{#2}}
\newcommand{\Nbar}{\bar{N}}

\begin{document}
\title{Limitations of tensor-network approaches for  optimization and sampling: \\ A comparison to quantum and classical Ising machines}
\author{Anna Maria Dziubyna}
\affiliation{Jagiellonian University, Institute of Theoretical Physics, \L{}ojasiewicza 11, 30-348 Krak\'ow, Poland}
\affiliation{Jagiellonian University, Doctoral School of Exact and Natural Sciences, \L{}ojasiewicza 11, 30-348 Krak\'ow, Poland}
\author{Tomasz \'{S}mierzchalski}
\affiliation{Institute of Theoretical and Applied Informatics, Polish Academy of Sciences, Ba{\l}tycka 5, 44-100 Gliwice, Poland}
\author{Bart\l{}omiej Gardas}
\email{bartek.gardas@gmail.com}
\affiliation{Institute of Theoretical and Applied Informatics, Polish Academy of Sciences, Ba{\l}tycka 5, 44-100 Gliwice, Poland}
\author{Marek M. Rams}
\email{marek.rams@uj.edu.pl}
\affiliation{Jagiellonian University, Institute of Theoretical Physics, \L{}ojasiewicza 11, 30-348 Krak\'ow, Poland}
\author{Masoud Mohseni}
\email{masoud.mohseni@hpe.com}
\affiliation{Emergent Machine Intelligence, Hewlett Packard Labs, CA, USA}

\begin{abstract}
Optimization problems pose challenges across various fields. In recent years, quantum annealers have emerged as a promising platform for tackling such challenges. To provide a new perspective, we develop a heuristic tensor-network-based algorithm to reveal the low-energy spectrum of Ising spin-glass systems with interaction graphs relevant to present-day quantum annealers. Our deterministic approach combines a branch-and-bound search strategy with an approximate calculation of marginals via tensor-network contractions. Its application to quasi-two-dimensional lattices with large unit cells of up to 24 spins, realized in current quantum annealing processors, requires a dedicated approach that utilizes sparse structures in the tensor network representation and GPU hardware acceleration. We benchmark our approach on random problems defined on Pegasus and Zephyr graphs with up to a few thousand spins, comparing it against the D-Wave Advantage quantum annealer and Simulated Bifurcation algorithm, with the latter representing an emerging class of classical Ising solvers. Apart from the quality of the best solutions, we compare the diversity of low-energy states sampled by all the solvers. For the biggest considered {i.i.d.} problems with over 5000 spins, the state-of-the-art tensor network approaches lead to solutions that are $0.1\%$ to $1\%$ worse than the best solutions obtained by Ising machines while being two orders of magnitude slower. We attribute those results to approximate contraction failures. {For embedded tile planting instances, our approach gets to approximately $0.1\%$ from the planted ground state, a factor of $3$ better than the Ising solvers.} While all three methods can output diverse low-energy solutions, e.g., differing by at least a quarter of spins with energy error below $1\%$, our deterministic branch-and-bound approach finds sets of a few such states at most. On the other hand, both Ising machines prove capable of sampling sets of thousands of such solutions.
\end{abstract}
\maketitle

\section{Introduction}
 
Ising spin-glass ground-state identification is a prime example of NP-hard problem~\cite{barahona_on-the-computational_1982} and a representative example of a discrete optimization problem~\cite{lucas_ising_2014}, where already a two-dimensional Ising system with fields provides a universal computational model~\cite{de_las_cuevas_simple_2016}.

Many physics-inspired approaches have been developed to tackle that problem. A prominent and general family of methods, starting with the Simulated Annealing algorithm~\cite{kirkpatrick_optimization_1983}, employs thermal fluctuations introduced by Metropolis-Hastings updates~\cite{metropolis_monte_1949, hastings_monte_1970} within a Markov-chain Monte Carlo (MCMC) paradigm. Rugged energy landscapes pose significant challenges for local Monte-Carlo updates, motivating elaborate schemes to navigate distinct local minima. Those include Parallel Tempering (PT)~\cite{earl_parallel_tempering_2005} to enhance thermal fluctuations, or heuristic cluster moves, like Iso-energetic or Houdayer cluster move~\cite{houdayer_cluster_2001, Zhu_efficient_2015}, or Nonequilibrium Monte Carlo with inhomogeneous temperature distribution over identified surrogate problem backbones~\cite{mohseni_nonequilibrium_2021}. {Alternative approaches to overcoming limitations of local updates include subgraph-based methods, which leverage fixed-treewidth substructures to improve sampling efficiency~\cite{selby_efficient_2014}. Additionally, highly optimized implementations of simulated annealing for Ising spin glasses have been demonstrated to achieve significant performance gains through multi-spin coding, particularly targeting previous graph geometry of D-Wave quantum annelers, i.e., Chimera graphs~\cite{isakov_optimised_2015}. }

\begin{figure*}[t]
\includegraphics[width=0.99\textwidth]{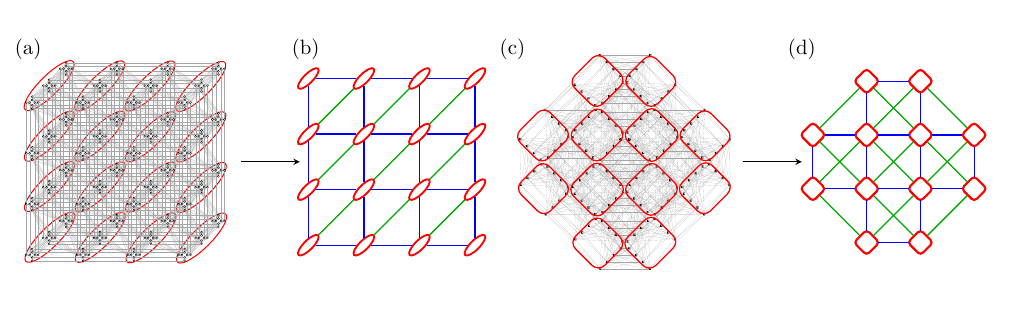}
\caption{Graph geometries appearing in the current D-Wave processors.
We show the interaction structures of the so-called Pegasus graph in panel (a), here with $384$ spins, and the Zephyr graph with $160$ spins in panel (c). Black dots denote spins, and couplings between spins are represented by grey lines. To apply the tensor network approach considered in this article, we employ quasi-2D structures of those graphs, with unit cells consisting of 24 spins for Pegasus and 16 spins for Zephyr, denoted by red lines in (a) and (c), respectively. We group the spins in each unit cell to arrive at a generalized Potts Hamiltonian in Eq.~\eqref{eq:factor_graph}, defined on a square lattice with the nearest neighbour (blue lines) and diagonal couplings (green lines) shown in panels (b) and~(d).
}
\label{fig:1}
\end{figure*}

An alternative approach under development employs quantum fluctuations within the quantum annealing paradigm~\cite{Kadowaki_quantum_1998, Santoro_theory_2002}. In particular, quantum annealers (QA) developed by D-Wave are a prime example of a quantum Ising machine, being a dedicated solver for Ising spin-glasses on quasi-two-dimensional (quasi-2D) geometries. A scaling advantage of quantum annealing over standard Monte Carlo methods remains a possibility based on analysis of some finite random problems~\cite{King_quantum_2023, bernaschi_quantum_2024, bauza_scaling_2024}. In idealistic QA, the time evolution is deterministic, with randomness entering while sampling from the final evolved state. Generally, there are no guarantees that this provides fair sampling from the low-energy manifold. For sufficiently fast quenches, D-Wave QA was shown to give results consistent with the coherent Schr\"{o}dinger dynamics~\cite{king_coherent_2022, king_computational_2024}. However, for slower quenches utilized for optimization in this article,  decoherence affects the outputs of the quantum simulator~\cite{bando_probing_2020}, adding to the randomness of the results.

Classical Ising machines include a recently developed Simulated Bifurcation Machine (SBM)~\cite{kashimata_efficient_2024, goto_combinatorial_2019, goto_highperformance_2021}. It maps the discrete Ising problem to a continuous system undergoing dissipative and chaotic dynamics governed by ordinary Hamilton equations. Stationary states of this process, after the final discretization of continuous variables, correspond to local minima of the original problem. The system evolves from a set of random initial states leading to multiple trajectories, simulations of which can be parallelized efficiently.

Tensor networks (TN) provide an alternative, typically deterministic, approach to studying the properties of many-body systems~\cite{verstraete_matrix_2008, orus_practical_2014, biamonte_tensor_2017, ran_tensor_2020, banuls_tensor_2023}.
In particular, they give a compact representation of the partition function of a classical low-dimensional spin system at arbitrary temperature~\cite{nishino_two-dimensional_2001} with information about marginal configuration probabilities following from (approximate) tensor network contractions. The underlying NP-hardness of the problem manifests itself in the difficulty of performing the contraction~\cite{schuch_computational_2007}. Nevertheless, we can turn TN into an optimization solver. 
In particular, Ref.~\cite{rams_approximate_2021} combined approximate tensor network contraction with a branch-and-bound approach, testing it in the context of the previous quantum-annealing processor geometry called the Chimera graph. The latter consists of unit cells of 8 spins forming a square lattice, a natural geometry for established approximate tensor-network contraction techniques. Therein, the TN-based approach was showing an advantage over PT and QA in some considered Chimera-geometry-native examples~\cite{rams_approximate_2021}.

At the same time, TN allows one to perform sampling~\cite{ueda_snapshot_2005}, which can be further enhanced in classical systems via combination with MCMC, where TN is used to provide non-local updates candidates~\cite{FriasPerez_collective_2023}.
Very recently, promising results have been reported~\cite{gangat_hyperoptimize_2024} for the application of hyperoptimized approximate contraction of tensor networks~\cite{gray_hyperoptimized_2024} to optimization problems with rugged energy landscape in two and three dimensions (using a simplified approach where optimization problem is addressed via direct sampling).

A promising alternative to working at finite temperature is provided by tropical TN~\cite{liu_tropical_2021}, considering the logarithm of the partition function in the zero-temperature limit. This is equivalent to a message-passing algorithm for ground state energy identification~\cite{mezard_information_2009}. However, automatic differentiation techniques applied in such TN contraction can also produce the corresponding ground state configurations. Nevertheless, this approach is currently limited to the exact contraction of the network and optimization problems with small treewidth.

This article explores the application of TN for graph geometries currently employed in D-Wave quantum anneals, which exhibit significantly enhanced connectivity over the previous generation processor and Chimera graph. For instance, such a graph can fit a cubic $15\times15\times12$ lattice of dimers~\cite{King_quantum_2023}. This poses significant challenges for employing approximate TN contraction techniques, which we tackle in this work. We aim to explore the limitations and gain insights into the sources of failure of our deterministic branch-and-bound TN approach, where increased problem complexity puts stringent constraints on the quality of the results outputted by our solver. We compare with representative examples of randomized algorithms, a class more typically employed in the context of optimization.

The rest of the article is organized as follows. In Sec.~\ref{sec:ising}, we discuss the Ising optimization problems, in particular, introducing the Pegasus and Zephyr graph geometries we are primarily interested in here. Sec.~\ref{sec:algorithm} provides a comprehensive description of our algorithm, including branch-and-bound search and the construction and contraction of a PEPS tensor network for graphs with large unit cells. We comment on exploiting sparse tensor structures and the variational zipper algorithm used in boundary MPS optimization. In Sec.~\ref{sec:LBP}, we discuss the optional concept of local dimensional reduction, employing loopy belief propagation in this context. In Sec.~\ref{sec:results}, we collect the results focusing on the ground state search and diversity of low-energy solutions. We benchmark our algorithm against D-Wave QA and SBM. We end with a discussion on how and why the tensor network approach is breaking in Sec.~\ref{sec:limitations}, followed by concluding remarks in Sec.~\ref{sec:discussion}. In the Appendix, we provide additional details on the stability of our approach, as well as other 2D graph geometries. Therein, we also discuss further details of the SBM algorithm.

\section{Ising problem}
\label{sec:ising}
We focus on the problem of finding the low--energy states of a classical Ising Hamiltonian~\cite{lucas_ising_2014},
\begin{equation}
H(\conf{s}{N}) =  \sum_{\langle i, j\rangle \in \mathcal{E}} J_{ij} s_i s_j + \sum_{i =1}^N h_i s_i,
\label{eq:Ising}
\end{equation}
where $\underline{s}_N$ denotes a particular configuration of $N$ binary variables $s_i=\pm 1$. The problem instance is defined by the set of real couplings $J_{ij}$ forming a graph $\mathcal{E}$ and local fields $h_i$. 
We will be using a notation where we denote a sub-configuration of the first $n$ variables as
\begin{equation}
\conf{s}{n} = (s_1, s_2, \ldots, s_n).
\label{eq:configuration}
\end{equation}

Here, we are interested in quasi-2D graphs. In particular, those of relevance for the current quantum annealing processors. D-Wave Systems has introduced two notable topologies in the realm of quantum annealing processors, named Pegasus and Zephyr, see Fig.~\ref{fig:1} panels (a) and (c), respectively. Pegasus and Zephyr represent a significant advancement over previous Chimera topology, featuring a quasi-2D lattice structure with improved connectivity and aiming to reduce noise~\cite{dattani_pegasus_2019, boothby_next-generation_2020, boothby_zephyr_2021}. In particular, the available Pegasus quantum processor, which we used as a reference, has 5616 qubits connected by over $40000$ couplers, where a typical qubit is connected to 15 different qubits. The Zephyr topology prototype we employed in this work has 563 spins,  a typical qubit connected to 20 different qubits, with the plan to expand the size of the chip to approximately $7000$ qubits. We provide further details on quantum annealing experiments in the Appendix.

\section{Tensor network approach}
\label{sec:algorithm}
To address graph geometries with large unit cells, such as Pegasus and Zephyr, using the tensor-network-based approach, we start by representing the Ising problem in Eq.~\eqref{eq:Ising} as a generalized Potts Hamiltonian with a reduced number of variables of larger dimensions. To that end, we group binary variables that form natural unit cells into clusters, as depicted in Fig.~\ref{fig:1} with red shapes. We group sets of $24$ binary variables for Pegasus geometry and $16$ variables for Zephyr geometry, resulting in generalized Potts Hamiltonian, 
\begin{equation}
    H(\conf{x}{\Nbar}) = \sum_{\langle m,n\rangle \in \mathcal{F}} E_{x_m x_n} + \sum_{n=1}^{\Nbar} E_{x_n}.
\label{eq:factor_graph}
\end{equation}
Here, $\mathcal{F}$ is a 2D square lattice with $\Nbar$ nodes; see Fig.~\ref{fig:1}(b) and (d) where we indicate nearest-neighbour interactions with blue lines and diagonal connections with green lines.
Each variable $x_n$ takes up to $d$ values with $d=2^{24}$ for Pegasus and $2^{16}$ for Zephyr geometry (in the maximal case when all qubits in the cluster are operational and are employed in a particular Ising Hamiltonian of interest). $E_{x_n}$ in Eq.~\eqref{eq:factor_graph} indicates an intra--node energy of the corresponding binary-variables configuration, and $E_{x_m x_n}$ is inter--node coupling. Our Potts Hamiltonian can also be considered as a factor graph~\cite{mezard_information_2009}, with interaction nodes limited to one-body and two-body terms, and $x_n$ constituting variable nodes.

\begin{figure}[t]
\begin{centering}
\includegraphics[width=0.99\columnwidth]{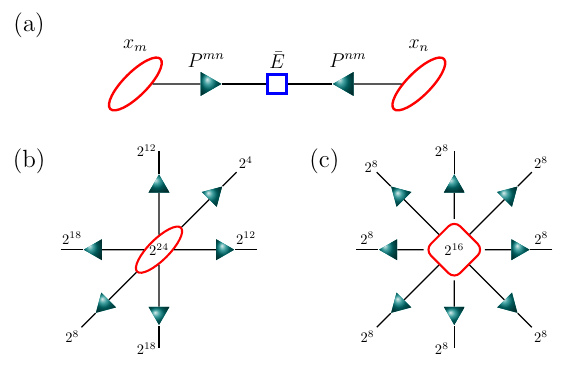}
\par\end{centering}
\caption{Compressed interaction matrices in clustered Hamiltonian. 
The number of degrees of freedom in each cluster is, respectively, $2^{24}$ for Pegasus geometry and $2^{16}$ for Zephyr geometry, which translates to linear dimensions of $E_{x_m x_n}$ matrices in Eq.~\eqref{eq:factor_graph}. However, they can be significantly reduced by employing the fact that there are spins in a given cluster that are not coupled to any spin in a neighbouring cluster. Projectors $P^{mn}$ and $P^{nm}$ project local degrees of freedom in each cluster to the relevant set of sub-configurations, limited to groups of interacting spins.
Following Eqs.~\eqref{eq:proj_energy} and \eqref{eq:projector}, this gives compressed interaction matrices $\overline E_{{\overline x}_{m}, {\overline x}_{n}}$, with linear dimensions for Pegasus and Zephyr geometries shown in panels (b) and (c), respectively.
}
\label{fig:2}
\end{figure}

We note that not all the spins in each red group in Figs.~\ref{fig:1}(a) and (c) interact with spins in neighbouring groups. It allows us to further compress the inter--node energy matrices, 
\begin{equation}
    E_{x_m x_n} = \bar{E}_{\bar{x}_m\bar{x}_n},
    \label{eq:proj_energy}
\end{equation}
where we introduce the projected variable
\begin{equation}
    \bar{x}_m = P^{mn}(x_m)
    \label{eq:projector},
\end{equation}
with $P^{mn}$ projecting $d$ configurations in $m$-th node onto $d^{mn}$ unique sub--configurations for a subset of original binary-variables that are actually coupled with spins in the $n$--th node. Similarly, in the opposite direction, $\bar{x}_n = P^{nm}(x_n)$. This is depicted in Fig.~\ref{fig:2}, together with the resulting dimensions.

\subsection{Tensor networks for optimization problems on quasi-2D graphs}

\begin{figure*}[b]
\begin{centering}
\includegraphics[width=\textwidth]{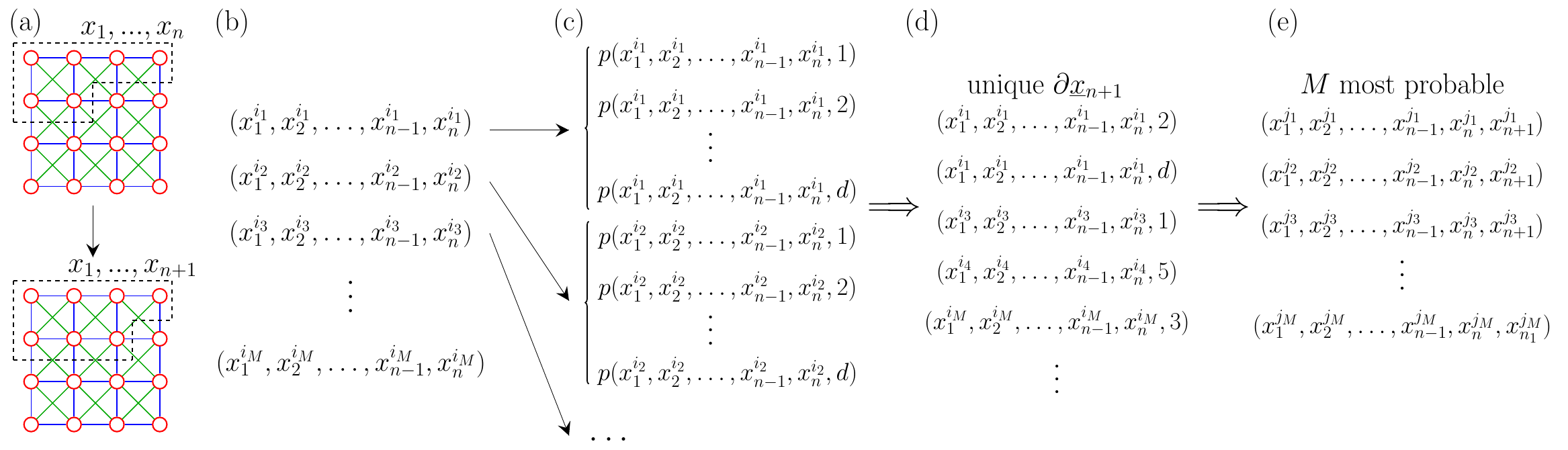}
\par\end{centering}
\caption{Branch and bound algorithm.
We translate the problem of identification of the low-energy configurations to the problem of finding the most probable configurations according to a Boltzmann distribution at some inverse temperature $\beta$, see Eq.~\eqref{eq:boltzmann}.
We sweep the 2D system processing row after row. At each step, we identify a set of $M$ configurations for $n$ clusters, $\conf{x}{n}$, with large marginal probabilities.
In a single step of the sweep, depicted in panel (a), we use $M$ configurations $\conf{x}{n}$ to identify $M$ configurations $\conf{x}{n+1}$, supported on one extra cluster. To that end, we branch $\conf{x}{n}$ configurations in panel (b) to include all possible values in $(n+1)$-th cluster, see panel (c), and calculate their marginal probabilities $p(\conf{x}{n+1})$. These marginal probabilities are determined from the approximate contraction of the PEPS tensor network representing the Boltzmann distribution, illustrated in Fig.~\ref{fig:6}. Next, we bound the set of candidate configurations in two steps. First, in panel (d), we employ locality of interactions in a 2D lattice, grouping the states according to their sub-configuration, $\partial\conf{x}{n+1}$, supported on a boundary with the remaining part of the lattice, i.e., with clusters $n+2, n+3, \ldots$. Only the most probable configuration for each unique boundary is retained for further processing. Discarded configurations encode excitation in the system and can be book-kept to map the low-energy spectrum of the system, as depicted in Fig.~\ref{fig:4}. Finally, in panel (e), we retain $M$ most probable configurations $\conf{x}{n+1}$, completing a single step of the sweep.
}
\label{fig:3}
\end{figure*}

We approach the problem of identifying low-energy states---from $2^N$ possible spin configurations---by translating it into the task of finding the most probable configurations according to a Boltzmann distribution at an inverse temperature $\beta$,
\begin{equation}
    p(\conf{x}{\Nbar}) = \frac{1}{Z} \exp{(-\beta H(\conf{x}{\Nbar}))},
    \label{eq:boltzmann}
\end{equation}
where $Z$ is a partition function. 
To effectively distinguish these low-energy states, we employ a branch and bound strategy, as illustrated in Fig.~\ref{fig:3}.

The strategy involves sweeping through the system to systematically build a set of high-probability configurations, adding one Potts variable at a time. At each step, we limit the number of partial configurations considered to at most $M$, keeping the computational complexity of the exploration process under control. This involves branching a set of $M$ configurations supported on the first $n$ clusters,  $\underline{x}_{n}$, to include possible configurations in $(n+1)$-th cluster, resulting in $d \cdot M$ trial configurations. For the subsequent step, we select $M$ of those with the largest marginal probabilities,
\begin{equation}
    p(\underline{x}_{n+1}) = p(x_{n+1} | \underline{x}_{n}) \cdot p(\underline{x}_{n}),
   \label{eq:chain_rule}
\end{equation}
which are calculated employing the chain rule. Here, we assume that variable indexing is consistent with the order in which the nodes are explored during the search (which can always be obtained with re-indexing).

\begin{figure*}[t]
\begin{centering}
\includegraphics[width=\textwidth]{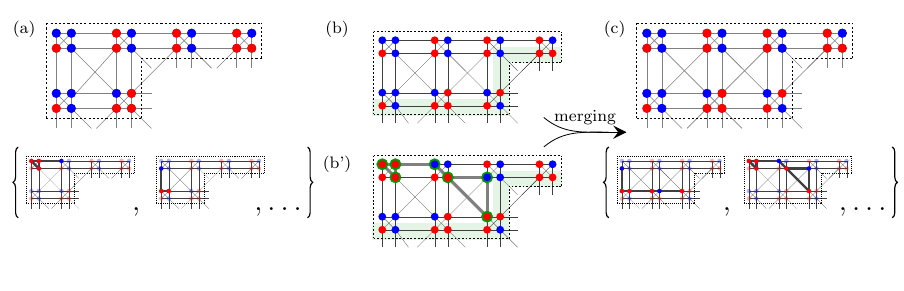}
\par\end{centering}
\caption{Excited states in the branch and bound algorithm. In each step of the sweep, we bound the set of candidate partial configurations, keeping the most probable ones for each considered {\it unique} boundary. We depict a single such configuration in the upper part of panel~(a) where dots represent spins: a blue dot denotes a spin up and red a spin down. It is accompanied by a set of low-energy excitations, shown in the lower part of~(a), that do not modify the boundary spins. In (b) and (b'), we show the merging process [step (d) in Fig.~\ref{fig:3}] of two spin configurations with the same boundary---highlighted in green. The one with a larger marginal probability becomes the main branch, depicted in the upper part of (c). The difference, indicated by bold grey lines in (b'), is added to the set of excitations above the main branch [lower part of (c)], together with excitations above (b) and (b') accumulated in the previous steps.
To keep their number under control, we keep only those that are singly connected and with a hamming distance between any pair of them above a specified threshold.
}
\label{fig:4}
\end{figure*}

The efficiency of the exploration is greatly enhanced, keeping $M$ fixed by exploiting the locality of the interactions. Namely, that the probabilities conditioned on partial configurations $\conf{x}{n}$ depend solely on the border part of $\conf{x}{n}$ directly interacting with the rest of the lattice,
\begin{equation}
\partial \underline{x}_{n} = \left \{ P^{kl}(x_k) : \langle k,l\rangle \in \mathcal{F}, k \le n, l > n \right \}.
    \label{eq:border}
\end{equation}
As such, we identify equivalent partial configurations $\conf{x}{n}$, which look the same from the point of the remaining part of the lattice. Specifically, for a given configuration $\underline{x}_{n}$ the conditional probability appearing in Eq.~\eqref{eq:chain_rule} relies solely on the subconfiguration at the border~$\partial \underline{x}_{n}$,
\begin{equation}
     p(x_{n+1} | \underline{x}_{n}) = p(x_{n+1} | \partial \underline{x}_{n}).
\end{equation}
This is used to merge configurations $\underline{x}_{n}$ with identical boundary subconfigurations $\partial \underline{x}_{n}$, see Fig.~\ref{fig:3}(d), 
which results in effective compression of the low energy manifold and improved performance of the algorithm. At the same time, the merging process enables us to identify low-energy excitation in the system as it reveals the most probable sub-configuration (with the lowest energy for fixed boundary sub-configuration), treated as the main branch, and low-energy excitations (spin glass droplets) above this partial ground states, see Fig.~\ref{fig:4}. 

The branch and bound procedure is applied iteratively until the last cluster, resulting in the ground state candidate and, from the merging process, a set of excitations that give an insight into the geometry of a low-energy manifold. 

We also keep track of the largest discarded probability during the optimization process $p_d$, which gives an upper bound for the probability of any discarded configuration. Thus, if the highest retained probability, corresponding to the state with the lowest identified energy, is larger than $p_d$, it would be sufficient evidence for ground state certification (for exact calculation of conditional probabilities; it remains a heuristic evidence for approximate tensor network contractions). In practice, reaching this condition proves to be infeasible in our examples, as large inverse temperature $\beta$ that facilitates such a situation makes approximate tensor network contraction numerically unstable. Still, intermediate $\beta$'s where approximate contraction still works reasonably well is sufficient for obtaining good quality solutions in the considered examples. We discuss those aspects in the following sections. 

\subsection{Construction and contraction of PEPS for graphs with large unit cells}

\begin{figure}[t]
\begin{centering}
\includegraphics[width=0.99\columnwidth]{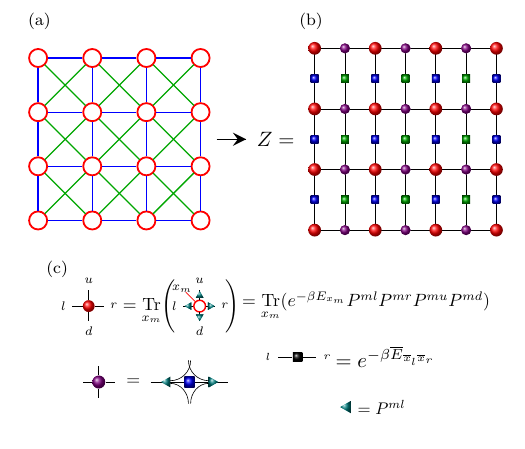}
\par\end{centering}
\caption{Representation of the partition function as a tensor network.
The partition function of a generalized Potts Hamiltonian in Eq.~\eqref{eq:factor_graph} on a square lattice with nearest neighbour and diagonal interactions, panel (a), can be expressed as a contraction of a tensor network depicted in panel (b), with individual tensors defined in (c).
Here, red dots denote site tensors, corresponding to clusters in (a), with 4 virtual legs mediating interaction structure and traced out Hamiltonian degrees of freedom.
Square marks denote exponents of corresponding interaction matrices, purple dots are mediating tensors that carry diagonal interactions, and triangles depict configuration projectors in Eq.~\eqref{eq:projector}. The result is a PEPS tensor network on a square grid, panel (b), that is amenable to established tensor network contraction strategies, see Fig.~\ref{fig:6}.}
\label{fig:5}
\end{figure}

To employ the algorithm described in the previous section, it is crucial to effectively compute the conditional probabilities $p(x_{n+1} | \partial \underline{x}_{n})$ to apply the chain rule in Eq.~\eqref{eq:chain_rule}. This can be achieved by representing all probabilities using a two-dimensional Projected Entangled Pair State (PEPS) tensor network~\cite{nishino_two-dimensional_2001, verstraete_criticality_2006}. For a classical system in Eq.~\eqref{eq:factor_graph}, the construction of a thermal state is exact and equivalent to the representation of its partition function~\cite{nishino_two-dimensional_2001, verstraete_criticality_2006},
\begin{equation}
Z = \sum_{\conf{x}{\Nbar}} \exp{(-\beta H(\conf{x}{\Nbar}))},
    \label{eq:partition_function}
\end{equation}
which is depicted in Fig.~\ref{fig:5}. Apart from {\it site tensors} carrying information about local energies and its interaction structure with neighboring sites (red circles in Fig.~\ref{fig:5}) and {\it bond tensors} encoding two-site interaction energies (squares), we introduce {\it mediating tensors} (purple circles). The latter encodes diagonal interactions in a tensor network defined on a square grid with nearest-neighbor connections only. The latter is viable to standard approximate tensor network contraction techniques.

\begin{figure}[t!]
\begin{centering}
\includegraphics[width=0.99\columnwidth]{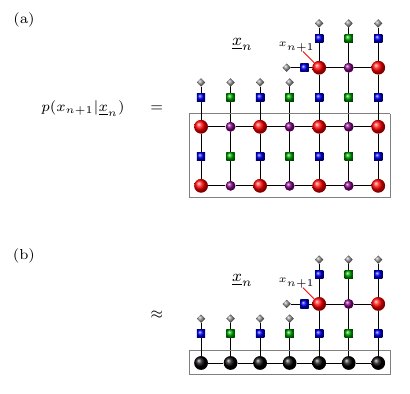}
\par\end{centering}
\caption{Calculation of conditional probabilities via approximate contraction of PEPS tensor network.
The calculation of marginal probabilities of interest boils down to the calculation of conditional probabilities $p(x_{n+1}|{\partial \conf{x}{n})}$ which, up to normalization, are represented by a tensor network in panel (a)---closely related to the one in Fig.~\ref{fig:5}(b). Here, the grey diamonds project the virtual degrees of freedom on a boundary $\partial \conf{x}{n}$ of a given configuration $\conf{x}{n}$.
A $(n+1)$th site tensor has an extra leg representing (un-traced) physical degrees of freedom. In panel (b), we systematically approximate the lower half of the network using the boundary MPS approach, see Fig.~\ref{fig:8}, where we mark the corresponding parts of (a) and (b) with grey boxes. This preprocessing step is done ones at the beginning of the calculation, being independent of a specific configuration $\conf{x}{n}$. Finally, the approximate network in (b) is contracted numerically exactly to retrieve the desired conditional probabilities.} 
\label{fig:6}
\end{figure}

\begin{figure}[t]
\begin{centering}
\includegraphics[width=0.99\columnwidth]{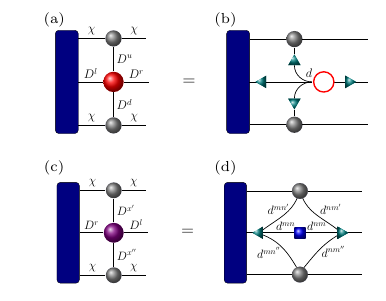}
\par\end{centering}
\caption{Sparse structure of PEPS tensor. Huge bond dimensions of the site and mediating tensors, which appear in the representation of Pegasus and Zephyr geometries, make their direct construction infeasible.
We employ the fact that in all calculations, i.e., in the randomized SVD, variational optimization in Fig.~\ref{fig:8}, and contractions in Fig.~\ref{fig:6}, they appear as part of diagrams such as the ones in~(a) and~(c). In all those diagrams, the constituent tensor and the result of their contraction each have only one leg, corresponding to the (huge) legs of the PEPS tensors. For the bond dimensions resulting from Pegasus and Zephyr graphs, those diagrams can still be contracted by employing the sparse structure of PEPS tensors and GPU hardware acceleration. For instance, in panel (b), the blue tensor and MPS grey tensors are projected [triangles are projectors in Fig.~\ref{fig:2}(a)] to a common physical space of a site tensor.
Here, they can be efficiently contracted along the connecting legs of dimensions $\chi$, and local Boltzmann factors (red circle) can be applied. Finally, the result is projected to the remaining virtual leg of the PEPS tensor (outgoing triangle).
Similar manipulations are performed in (d) for optimal contraction of diagrams involving mediating tensors.}
\label{fig:7}
\end{figure}

Finding the ground state of a general Ising spin glass problem is NP-hard~\cite{barahona_on-the-computational_1982}. In the tensor network representation of spin glass problems, this hardness manifests itself in the PEPS tensor network being $\# P$-hard to contract exactly~\cite{schuch_computational_2007} (extract information from). Higher than NP-hard computational class reflects the fact that it contains information about the entire spectrum, not only the ground state. The exponential computational cost for exact contraction motivates using approximate (heuristic) contraction schemes, reducing the cost to polynomial in lattice parameters.

In this work, we adopt boundary matrix product states (MPS) approach~\cite{verstraete_matrix_2008, lubasch_unifying_2014}, illustrated in Fig.~\ref{fig:6} in the context of tensor network representation of the conditional probability in Eq.~\eqref{eq:chain_rule}. Tensors in the gray box in Fig.~\ref{fig:6}(a) are approximated in Fig.~\ref{fig:6}(b) by a boundary MPS (black dots)
with bond dimensions truncated to $\chi$. For lattices with multiple rows, boundary MPS approximations are obtained sequentially row after row. In each step, a product of boundary MPS from the previous row and
the transfer matrix formed by the next row of tensors, having the structure of matrix product operator (MPO), is approximated by a new boundary MPS of limited bond dimension $\chi$. We describe the procedure we use in detail in Sec.~\ref{sec:zipper}. This calculation of boundary MPSs is done once as a preprocessing step of the algorithm, as they are shared by all $p(x_{n+1}| \partial{\underline{x}}_{n})$ of interest. Finally, diagrams like in Fig.~\ref{fig:6}(b) are viable for numerically exact contraction.
In this way, information from the entire lattice is included in the calculation of probabilities.

Execution of the above techniques for tensor sizes needed to encode the Ising Hamiltonian relevant to near-term quantum annealing technology requires a dedicated approach, which we describe below.

\begin{figure*}[t!]
\begin{centering}
\includegraphics[width=0.95\textwidth]{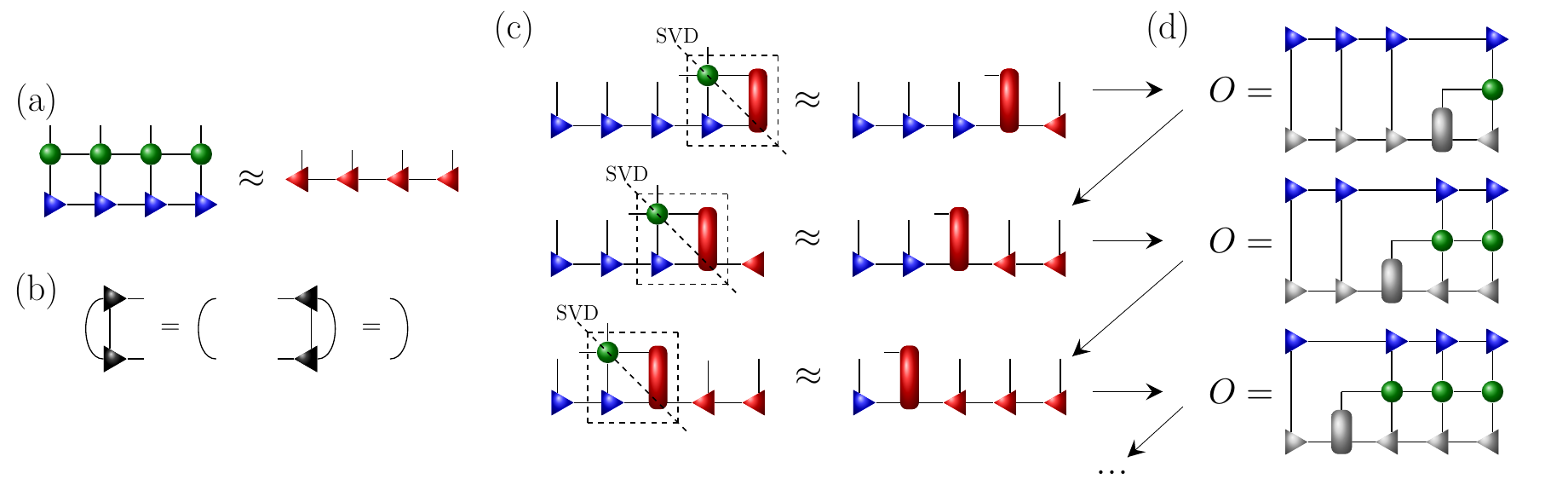}
\par\end{centering}
\caption{Systematic approximation of the lower half of PEPS lattice via boundary MPS. In panel (a), a boundary MPS for $(k+1)$-th row with bond dimension $\chi$ (blue triangles) is contracted with an MPO representing the $k$-th row of PEPS tensors (green dots), and approximated by a new boundary MPS of bond dimension $\chi$ for $k$-th row (red triangles). To that end, we employ a zipper scheme, where PEPS tensors are attached sequentially, one at a time, from right to left. Panel (c) starts with the initial MPS in the left canonical form [right(left)-pointing triangles represent left(right)-canonical MPS tensors, see (b)]. The first MPO tensor is attached, and a tensor in the dashed box is approximated via SVD, retaining $\chi$ leading singular values. Here, the initial connecting tensor, represented by a red rectangle, is a trivial unity with all bond dimensions equal one. The procedure is repeated in subsequent sites, with the MPS tensors to the left (right) of the connecting tensor in right (left) canonical form. This mixed canonical form makes an SVD approximation locally optimal. The problems we consider in this article have huge MPO bond dimensions, which requires resorting to randomized truncated SVD~\cite{halko_finding_2009, martinsson_randomized_2020}. To increase the numerical stability, each step of a zipper procedure is followed by a variational optimization~\cite{verstraete_matrix_2008} of a new MPS, maximizing the overlaps in panel~(d). 
}
\label{fig:8}
\end{figure*}

\subsubsection{Employing sparsity of tensors}
Sizes of tensors appearing in the TN representation of marginal probabilities for Pegasus and Zephyr geometries prevent their direct construction. Specifically, for a {\it site} tensor the bond dimensions $D^l, D^u, D^r$, $D^d$, see Fig.~\ref{fig:7}(a), in a full Pegasus graph are $2^{18}, 2^{14}, 2^{14}, 2^{18}$, respectively, while in a full Zephyr graph they read $2^{16}$, $2^8$, $2^8$, $2^{16}$ 
(the values might be permuted depending on the direction from which the boundary MPS contraction of a 2D lattice is done). Calculation of a diagram in Fig.~\ref{fig:7}(a), which is the building block of all contraction algorithms that we use, would scale as $\mathcal O(D^{l} D^{u} D^{r} D^{d} \chi^2)$ if the site tensor was constructed directly, making such an approach infeasible in practice.

A strategy to address this challenge is to leverage the sparse structure of this tensor, shown in Fig.~\ref{fig:5}(c). We employ it in Fig.~\ref{fig:7}(b), which depicts an efficient strategy to contract the diagram. First, the legs of the tensors, to be contracted with the legs of the site tensor, are projected back to the physical space of the site tensor of size $d$ (triangles). Virtual legs of those tensors are next contracted using batched matrix multiplication at a numerical cost of $\mathcal O(\chi^3 d)$. Additionally, the latter greatly benefits from hardware acceleration on GPU. Still, it reminds the computational bottleneck for the Pegasus graph with $d=2^{24}$, which, together with available memory, limits the feasible boundary MPS bond dimensions to approximately $\chi=10$~\cite{GPU}. Finally, the local energy term is applied, and the physical space is projected by the last right-pointing projector in Fig.~\ref{fig:7}(b), completing the contraction of the diagram.

A similar approach is needed for operations involving mediating tensor, which dimensions $D^r, D^{x'}, D^l$, $D^{x''}$, see Fig.~\ref{fig:7}(c), are $2^{14}$, $2^4$, $2^{18}$, $2^8$ for a full Pegasus graph, and all equal to $2^{16}$ for the Zephyr graph. The optimal contraction scheme employs the internal structure of this tensor, shown in Fig.~\ref{fig:7}(d), where bond dimensions $d^{mn'}, d^{mn}, d^{mn''}, d^{nm'}, d^{nm}, d^{nm''}$ correspond to the dimensions in Fig.~\ref{fig:2}(b) and (c). For Pegasus graph, they read $2^4, 2^{12}, 2^0, 2^0, 2^{18}, 2^8$, and for Zephyr all are equal to $2^8$. The specific order in which projectors, boundary tensors, and bond tensors get combined in Fig.~\ref{fig:7}(d) depends on specific values of those dimensions and, hence, also on a graph/diagram rotation. For a full Pegasus graph, the number of numerical operations is $\mathcal O(2^{26} \chi^3 )$. For a Zephyr graph, it scales as $\mathcal O(2^{32} \chi^3)$, being a computational bottleneck for this lattice. Those operations again benefit from GPU hardware acceleration, though available memory on GPU and the size of intermediary tensors appearing during the contraction again limit feasible $\chi$ to around $10$~\cite{GPU}. 

\subsubsection{Variational zipper algorithm for boundary MPS}
\label{sec:zipper}
Construction of boundary MPS, which we employ in approximate PEPS contraction, requires systematic approximation of MPO-MPS product by an MPS with a limited bond dimension $\chi$. A particular challenge here is the bond dimensions of MPO tensors ({\it site} and {\it mediating} tensors); see the previous section.

To that end, we combine a zipper scheme of Ref.~\cite{sinha_efficient_2024} with the standard variational optimization of the resulting MPS that aims to iteratively maximize its overlap with the target MPO-MPS product~\cite{verstraete_matrix_2008}.
The procedure is depicted in Fig.~\ref{fig:8}. The zipper scheme applies MPO tensors one at a time while gradually switching the canonical form of the resulting MPS. The latter makes each SVD approximation of a tensor represented by dashed boxes in Fig.~\ref{fig:8}(c), used to truncate the bond dimension, globally optimal at a given step of the zipper.
To handle MPO tensors bond dimensions related to Pegasus and Zephyr geometries, we utilize a truncated randomized SVD scheme~\cite{halko_finding_2009, martinsson_randomized_2020} that directly targets the desired number of dominant singular values and vectors of interest. It operates in a matrix-free way, i.e., it does not require an explicit construction of the decomposed tensor, only a procedure calculating its multiplication with a trial vector from the left and right sites. Those take the form of diagrams such as in Fig.~\ref{fig:7}, allowing us to utilize the sparse structures of MPO tensors.

Randomized SVD algorithms targeting a subspace of singular values, while much faster and memory efficient than the full-rank SVD algorithms based on Householder reflections~\cite{golub_matrix_2013}, are numerically unstable and prone to getting stuck in local optima. To mitigate this issue, we do the truncation in two steps, targeting $2 \chi$ singular values with randomized SVD, which, after variational refinement, are further truncated to the target $\chi$ singular values. More importantly, each step of the zipper procedure is followed by a sweep of variational optimization of the instantaneous stage of the zipper procedure, as shown in Fig.~\ref{fig:8}(d). To limit the computational effort, we can limit the range of the variational sweep to a few tensors in the vicinity of the instantaneous centre of the zipper. 

We have found the above procedure to be most stable, allowing us to operate with $\chi$ up to approximately $10$ for the targeted lattices with the available resources~\cite{GPU}. We have also tested a strategy where a full zipper sweep was performed first to give the initial state for subsequent variational optimization, as well as a strategy where only the variational optimization was used in combination with a gradual increase of inverse temperature $\beta$. Both those approaches, however, were resulting in worse performance.

\begin{figure}[t]
\begin{centering}
\includegraphics[width=0.99\columnwidth]{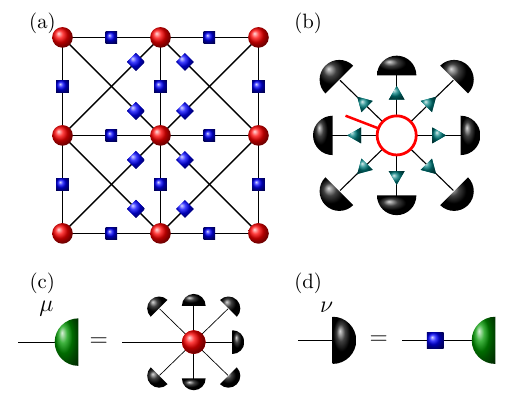}
\par\end{centering}
\caption{Local dimension reduction. To lower the numerical cost of tensor network contraction, we employ a loopy belief propagation (LBP) algorithm to approximate the marginal probabilities of each cluster variable and retain a given number of the most probable (according to LBP) configurations in each cluster. This substantially reduces the cost of subsequent global branch and bound search. An LBP can be rephrased as a simple approximate contraction scheme of a tensor network representing the partition function, such as in panel (a) [see Fig.~\ref{fig:5}(a)], aiming at an approximation of local marginals in panel (b). Here, black half-circles are boundary vectors (messages from edges to clusters), approximating the influence of the rest of the lattice. They are obtained via an iterative procedure shown in panels (c) and (d), where one sequentially updates, in panel (c), messages from clusters to edges [boundary tensors represented by green half-circles], and, in panel (d), messages from edges to clusters. In our numerical experiments, we perform a fixed number of such updates before the estimation of local marginal probabilities in panel (b).
}
\label{fig:9}
\end{figure}

\section{Local dimensional reduction}
\label{sec:LBP}
We additionally explore the combination of the algorithm in Sec.~\ref{sec:algorithm} with local dimension reduction of cluster degrees of freedom, which significantly reduces the numerical cost of subsequent tensor network contractions. It is done by selecting the most probable states in each cluster, where we approximate the marginal probabilities of each Potts variable employing Loopy Belief Propagation (LBP) algorithm~\cite{pearl_reverend_1982, yedidia_understanding_2003, mezard_information_2009}.

We follow the standard LBP formulation~\cite{mezard_information_2009}, incorporating within it the projectors in Eq.~\eqref{eq:projector} to significantly reduce the memory cost of LBP representation for problems with large local dimensions. The reduction is indispensable in addressing dimensions we encounter in Pegasus and Zephyr geometries after clustering.

Specifically, the LBP procedure aims at finding marginal probabilities $p(x_n)$ of configurations $x_n$ in a cluster $n$ by iteratively updating messages from clusters to edges and from edges to clusters. Such iterations converge to exact results for problems defined on a tree. Much worse performance is expected for lattices with loops, such as the ones we considered here. However, the results only serve to select a fraction of local degrees of freedom for each cluster, and as such, this approach is viable for testing. The main steps of the procedure are outlined in Fig.~\ref{fig:9} and can naturally be interpreted as a form of approximate tensor network contraction. 

Below, we simplify the LBP formulation for general factor graphs~\cite{mezard_information_2009}, which can contain interactions between arbitrary numbers of variables, to the case of at most two-site terms appearing in Eq.~\eqref{eq:factor_graph}. The messages from edges to clusters, $\nu_{mn}(\overline{x}_n)$, where clusters' indices $m$ and $n$ uniquely identify directional edges for two-site interactions, are depicted as black semicircles in the Fig.~\ref{fig:9}. They are initialized as a uniform distribution, $\nu_{mn}(\overline{x}_n) \sim 1$, implying equal probabilities for all states. The messages from the clusters towards the edges of the graph, which is denoted by $\mu_{mn}(\overline{x}_m)$, denoted with green semicircles in Fig.~\ref{fig:9}, follow from an update
\begin{equation}
     \mu_{mn}(\overline{x}_m){=}{\sum_{P^{mn}(x_m)=\overline{x}_m}}{\exp{(-\beta E_{x_m})}}{\prod_{m' \in N(m)\setminus n} \nu_{m' m}(x_m)},
\end{equation}
see Fig.~\ref{fig:9}(c). Above, $N(m)$ indicates sites interacting with site $m$, and we employ projectors in Eq.~\eqref{eq:projector} where $\nu_{m' m}(x_m) = \nu_{m' m}(\overline{x}_m)$, cf., the definition of {\it site tensor} represented as red circles in Fig.~\ref{fig:5}. Subsequently, one updates the message from edge to clusters 
\begin{equation}
    \nu_{mn}(\overline{x}_n)= \sum_{\overline{x}_m} \exp{(-\beta \overline{E}_{\overline{x}_n \overline{x}_m})} \mu_{mn}(\overline{x}_m), 
\end{equation}
as depicted in the Fig.~\ref{fig:9}(d).

After several such iterations (a fixed number or after reaching convergence), we can approximate the marginal probabilities of spin configurations associated with each cluster within the graph.
In particular, the probability of configuration $x_v$ is proportional to the product of all messages originating from neighboring edges linked to the respective cluster, see Fig.~\ref{fig:9}(b), 
\begin{equation}
    p(x_{n}) \propto  \exp{(-\beta E_{x_n})} \prod_{m \in N(n)} \nu_{m n}(x_n).
\end{equation}
For dimensional reduction, we select the desired number of most probable states in each cluster according to such a procedure.

{We note that within the temperature range considered in this study, BP messages do not fully converge with the maximal change of messages during iteration saturating at around $10^{-3}$. Hence, reaching saturation in convergence, messages after a predefined number of iterations are used to select representatives of each cluster. At sufficiently high temperatures (small $\beta$),  BP would be expected to converge. However, such temperature regimes are not practically relevant to the optimization problems studied here. Finally, we should note that the selection based on local marginals may not be able to identify local configurations contributing to a globally optimal state.}

\section{Results}
\label{sec:results}
In this section, we present three classes of instances that we used for testing the algorithms of Sec.~\ref{sec:algorithm} and~\ref{sec:LBP} and discuss the performance metrics we adopt. Finally, we show the results for problems defined on Pegasus and Zephyr geometries. Apart from the TN-based algorithm described Secs.~\ref{sec:algorithm} and~\ref{sec:LBP}, all instances have been executed on D-Wave quantum annealer and optimized using SBM algorithm~\cite{kashimata_efficient_2024, goto_combinatorial_2019, goto_highperformance_2021}. In addition, we collect further evidence in the Appendix, including problems defined on a simple square lattice and square lattice with diagonals. Therein, we also show the dependence of our algorithm's performance on various parameters, most notably the inverse temperature, $\beta$.

\subsection{Problem instances}
We employ random instances with all available couplings drawn from the same distribution, primarily focusing on Pegasus and Zephyr coupling graphs~\footnote{As some of the qubits and couplers that would form an ideal Pegasus/Zephyr lattice are not available in the quantum annealing hardware, we set the corresponding local fields and couplings to zero so that all solvers use the same instances}, see Fig.~\ref{fig:1}.
\begin{itemize}
\item class I. We employ uniform distribution in $[-1, 1]$ for $J_{ij}$ and set local fields $h_i = 0$, see Eq.~\eqref{eq:Ising}. Those instances have a global reflection symmetry.
\item class II: Similarly as in Class I, instances are defined by couplings drawn from uniform distribution in $[-1, 1]$. We include small local fields with $h_i$ drawn from an uniform distribution in $[-0.1, 0.1]$, breaking reflection symmetry.
\item class III: We adopt the Corrupted Biased Ferromagnets instances formulated on Pegasus graph (CBFM-P) of Ref.~\cite{tasseff_emerging_2022}. They are defined by discrete distributions with
$P(J_{ij} = 0) = 0.35$,  $P(J_{ij} = -1) = 0.10$, $P(J_{ij} = 1)=0.55$, and $P(h_i = 0) = 0.15$,  $P(h_i = -1) = 0.85$.

\item {class IV: We utilize tile planting instances on a square spin lattice, following the method outlined in Ref.~\cite{perera_computational_2020,perera_chook_2020}. The instances defined on a square lattices with the system sizes $L = 10$, $L=20$, $L=40$ are subsequently embedded into Pegasus graphs. After embedding, the number of spins increases to 164, 721 and 3065 spins respectively. The instances are constructed with a planted solution, meaning that the ground state is known a priori. The complexity of the planted instances is controlled by the probabilities $p_1$, $p_2$ and $p_3$, such that $p_1 + p_2 + p_3 \leq 1$. We set $p_2=1.0$ for which the instances are expected to be the hardest to solve.}

\end{itemize}
We summarize the considered problem classes in Tab.~\ref{tab:instances}.

The probability distribution defining instances in class III was selected for Pegasus geometry through a trial-and-error method to ensure that problems from this class are challenging for classical algorithms. Specifically, Ref.~\cite{tasseff_emerging_2022} presents extensive benchmarks comparing the time-to-solution between D-Wave Advantage quantum annealer and a selection of classical algorithms, including simulated annealing, parallel tempering with isoenergetic cluster moves, integer quadratic programming, message passing, etc. (but not SBM). It demonstrated that in this class, and for some range of target approximation ratios, quantum annealer provides run-time benefits over a tested collection of established classical methods. {A recent comparison with the SBM algorithm was performed in~\cite{vodeb_accuracy_2024} on max-cut problems showing the competitive performance of SBM.}

\begin{table}[t]
\begin{tabular}{c|c}  
Class & Description \\
\hline
\hline
  I & Random couplings in $[-1, 1]$. \\
\hline
  II & Class I + random local fields in $[-0.1, 0.1]$. \\
\hline
  III & CBFM-P instances of Ref.~\cite{tasseff_emerging_2022} for Pegasus graph. \\
  \hline
  IV & {Tile planting instances of Ref.~\cite{perera_computational_2020,perera_chook_2020} for Pegasus graph.}
\end{tabular}
\caption{Classes of problem instances considered in this article.}
\label{tab:instances}
\end{table}

\begin{figure*}[t]
\begin{centering}
\includegraphics[width=0.99\textwidth]{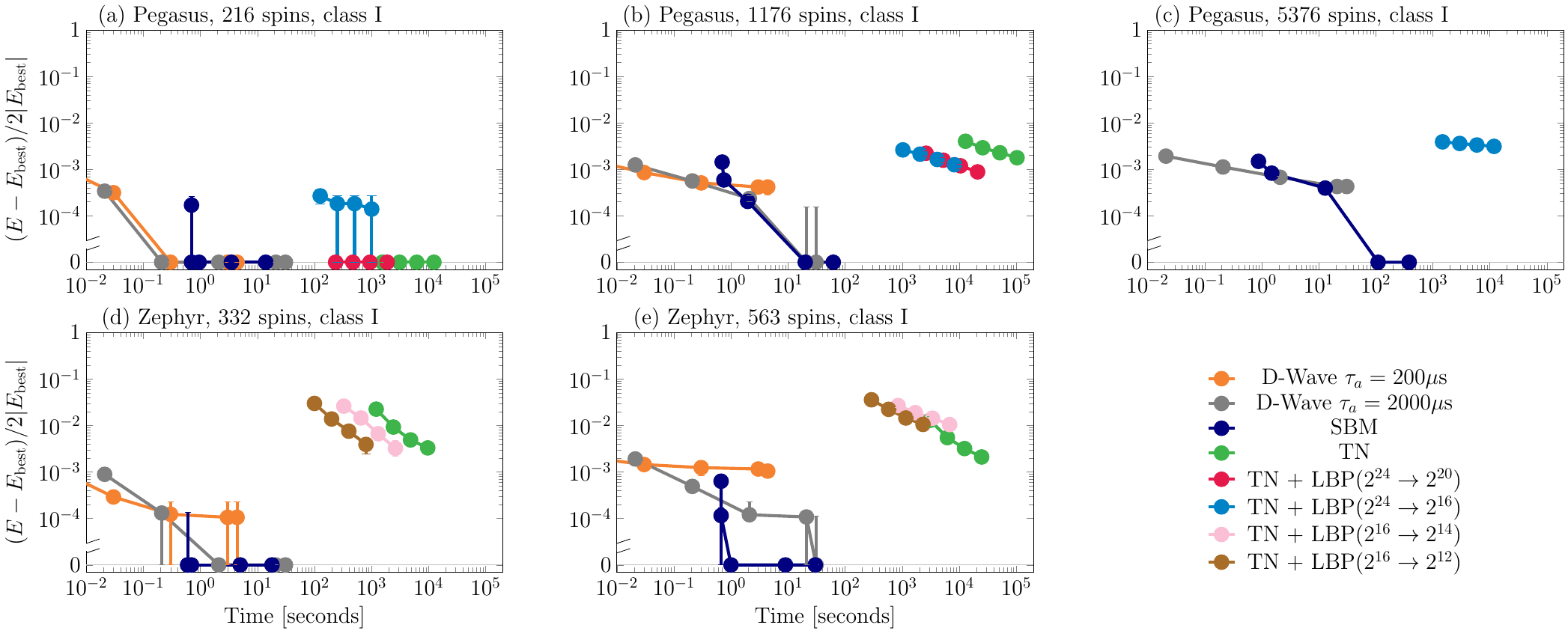}
\par\end{centering}
\caption{Time to approximation ratio. We focus on class I instances for Pegasus and Zephyr for a selection of system sizes. We compare the considered solvers in the time needed to reach a given quality of a solution quantified by energy above the best identified one selected among D-Wave QA (orange and grey), SBM (dark blue), and TNs. Here, we show the median based on 20 instances. Apart from full-space tensor-network results (green), we plot results obtained with dimensional reduction of local cluster degrees of freedom: truncation from $2^{24}$ states for Pegasus to $2^{16}$ states (blue) and $2^{20}$ states (red) and truncation from $2^{16}$ states for Zephyr to $2^{12}$ states (brown) and $2^{14}$ states (pink). The points for QA and SBM are from $10$, $100$, $1000$, $10000$ and all available samples (trajectories). Here, we estimate QA time as annealing time plus $100\mu s$. SBM and TN times are for a single GPU~\cite{GPU}. While the TN algorithm is in principle deterministic, it can be executed starting from different corners of a quasi-2D lattice (8 lattice transformations in total), and the points for TN correspond to trying $1$, $2$, $4$, and $8$ such transformations. Here, we used $\beta = 0.5$ for Pegasus and $\beta = 1.0$ for Zephyr instances, MPS bond dimension $\chi = 8$, the maximal number of branches in the search $M=1024$. The error bars of the median follow from bootstrapping.
}
\label{fig:10}
\end{figure*}

\subsection{Metrics: best energy, diversity of solutions}
\label{sec:metrics}
We use two fundamental performance metrics: the best energy found and the diversity of identified low-energy configurations, both within the framework of a time-to-solution. 

In the first metric, we focus on the lowest energy obtained with given resources, which we quantify using solver runtime. We will be plotting a relative energy difference above the best energy found, 
\begin{equation}
d_E = (E - E_{\mathrm{best}}) / 2 |E_{\mathrm{best}}|,
\end{equation}
where the reference energy $E_{\mathrm{best}}$ is established by collecting the results of all tested solvers {when the ground truth is unknown.}

In the second metric, we probe the quality of sampling of diverse approximate solutions of a spin-glass problem~\cite{mohseni_sampling_2023, zucca_diversity_2021}. We summarize the idea behind the diversity measure below. {We focus on good-quality solutions with energies within a given {\it approximation ratio} $a_r$, that is, the solutions for which $d_E \le a_r$.
The goal is to select from the available set of such solutions} the largest set of independent solutions. The size of the set provides a diversity of solutions. We refer to two spin configurations $\conf{s}{N}$ and $\conf{s'}{N}$, using here original spin variables, as independent if their Hamming distance $d$ is above a given threshold
\begin{equation}
    d(\conf{s}{N}, \conf{s'}{N}) \geq R N,
\label{eq:diversity}
\end{equation}
where $N$ is the number of spins and $R \in [0, 1]$ is the relative distance threshold. Indeed, a large threshold value makes it unlikely to transition between two such solutions by relying on local updates. Below, we set $R=1/4$ and $R=1/2$. For class I instances, we do not distinguish solutions differing by a global reflection, i.e., we first flip some solutions, to have the same value of the first spin in all samples.

\begin{figure*}[t!]
\begin{centering}
\includegraphics[width=0.95\textwidth]{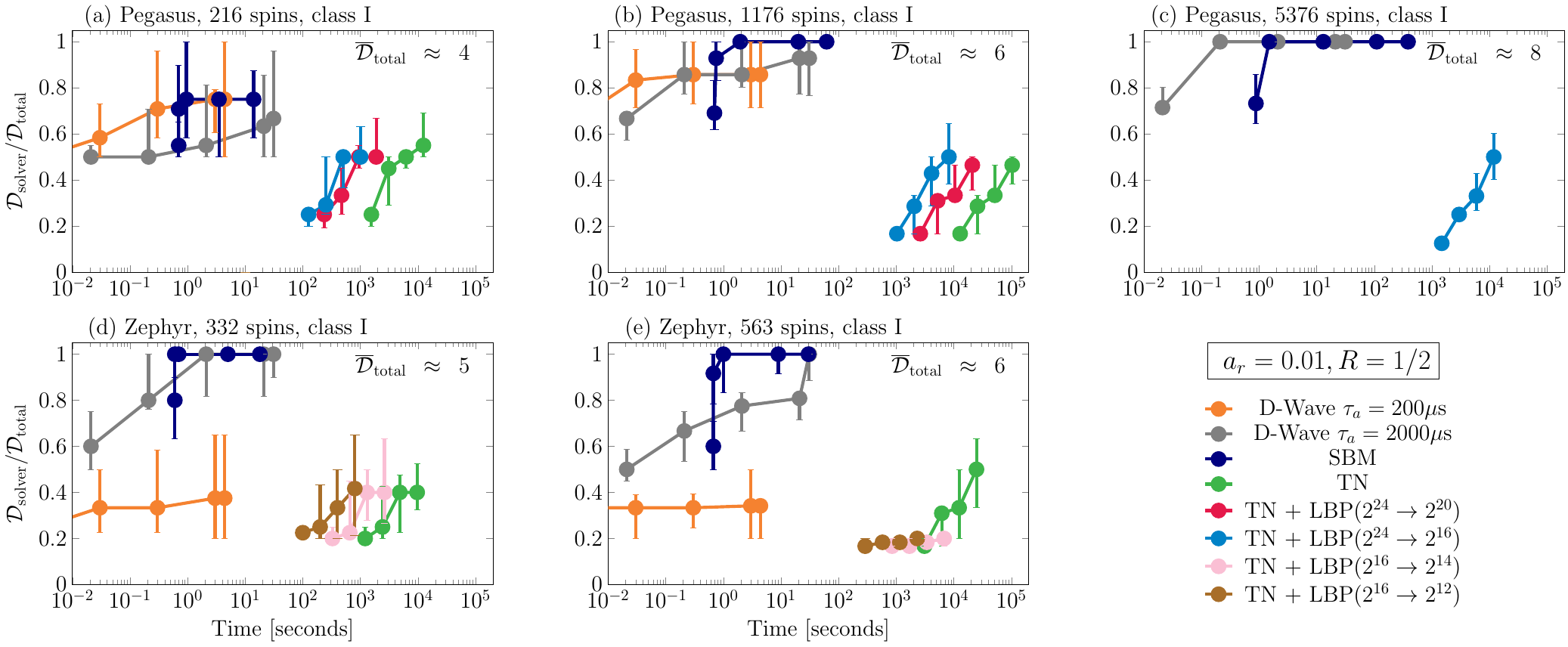}
\par\end{centering}
\caption{Time to diversity ratio. We show the time needed to reach a given median fraction of diverse solutions within $a_r=0.01$ for class I instances of the Pegasus and Zephyr graphs. The total diversity of each instance comes from combining the outputs of all considered solvers. The median among 20 instances is annotated, as $\overline{\mathcal{D}}_{\rm total}$ in each panel. We focus on the relative distance threshold $R=1/2$. It gives sets of independent solutions consisting of only a few spin configurations, and all $3$ solvers are able to identify such varying states. The simulation parameters are the same as in Fig.~\ref{fig:10}.
}
\label{fig:11}
\end{figure*}

To compare the performance of considered solvers, we proceed as follows.
We start with the estimation of the ground truth diversity $\mathcal{D}_{total}$ of a given problem instance. To that end, we collect the solutions from all solvers and search for the largest set of independent solutions within given approximation ratio $a_r$. This is equivalent to finding a maximum clique (i.e., a subgraph of the analyzed graph where each pair of nodes is adjacent) in the graph where low-energy solutions form the graph's nodes, and any two low-energy states with a Hamming distance greater than the threshold are adjacent. The maximum clique problem is known to be NP-complete~\cite{bomze_handbook_1999}, and we approximate its solution using a randomized heuristic approach~\footnote{We iterate over the list of states in a random order, building in the process a set of independent states. At each step of the iteration, a configuration is added to the list of independent states if it is independent of all configurations already in the list. We select the largest set obtained in all the restarts of that procedure. Typically, we observe saturation in obtained clique sizes after performing on the order of $100$ restarts.}. $\mathcal{D}_{total}$ is equal to the size of the set obtained in the above procedure. 

The obtained reference set of independent solutions is then used to calculate the diversity of solutions $\mathcal{D}$ found by each solver for a given problem instance. To that end, we iterate over all low-energy solutions obtained by a given solver and mark the reference independent solution to which they are the closest. The diversity for the solver is equal to the number of marked states.

\subsection{Results of the ground state search}

Figure~\ref{fig:10} shows the time to median approximation ratio for 20 instances of class I for Pegasus geometry with 216 spins, 1176 spins, and 5376 spins and Zephyr graph with 332 spins and 563 spins. Analogous data for other instance classes are collected in the Appendix{, where we observe similar performance for instances in classes I-III. For the largest considered problems in class IV, the TN algorithm outperforms QA and SBM in terms of the best energy found, improving $d_E$ by a factor of $3$ to $4$.}

Apart from full-space TN simulations, we additionally conduct simulations with local dimensional reduction of cluster degrees of freedom, as described in Sec.~\ref{sec:LBP}. Here, we cut the $2^{24}$ cluster states to $2^{20}$ and $2^{16}$ states for Pegasus instances, and from $2^{16}$ states to $2^{14}$ and $2^{12}$ states for Zephyr instances. Due to the memory limitations,
for Pegasus instances with 5376 spins, we only performed calculations with local dimensional reduction to $2^{16}$ states. The parameter selection was performed separately for each type of graph: an inverse temperature $\beta = 0.5$ was used for Pegasus instances, while $\beta = 1.0$ was used for Zephyr instances. We further show the influence of $\beta$ on the quality of the results in the following section and in Appendix~\ref{stability}.

For the smallest Pegasus graph in Fig.~\ref{fig:10}(a), TN simulations with no dimensional reduction find the ground state (for median instance). Employing dimensional reduction to $2^{16}$ prevents us from identifying the ground state in some cases -- we have checked that this is due to dimension reduction that no longer supports the ground state configuration, showing the limitations of the LBP preprocessing. For larger Pegasus lattices in Fig.~\ref{fig:10}(b) and (c), the best energies reached are systematically below the approximation ratio of $10^{-2}$ with and without invoking the dimensional reduction. We are not, however, typically able to find the ground state in those cases. The outcome of the TN procedure depends on the edge of the quasi-2D lattice from which the branch-and-bound is initiated, marked by the improvement of the best results after collecting data from various such trials. It indicates deficiencies in the stability of the approach for feasible execution parameters.

Zephyr geometry proves to be more challenging to our approach; see Fig.~\ref{fig:10}(d) and (e). In some cases, we observe a trade-off between TN contraction stability and the degree of dimension reduction. While overall, the best results are obtained for no reduction, there are instances where the best result, even the ground state, is identified only after preceding TN-based branch-and-bound search with a local dimensional reduction. 
We expect the drop in performance of our TN-based approach for Zephyr geometry, as compared to Pegasus, is due to the larger fraction of diagonal interactions in the former and its impact on contraction stability.
In the Appendix~\ref{sec:square}, we show the results for an Ising model defined on a plain square lattice and a square lattice with additional diagonal terms, where TN produce the best solutions.

Finally, Fig.~\ref{fig:10} includes the results of QA and SBM. Both these approaches 
operate on a timescale that is orders of magnitude smaller than TN, and
typically allow us to identify the ground state. The probabilistic nature of these algorithms makes the operation time roughly proportional to the number of samples, with $10^4$ samples typically sufficient to identify the ground state in our examples. In contrast, the TN algorithm is deterministic and requires completing a systematic sweep through the system to return all identified solutions in one go.

\subsection{Diversity of solutions}
Figure~\ref{fig:11} shows the time to median diversity ratio calculated for the same set of instances as in Fig.~\ref{fig:10}. Here, we focus on a large distance threshold $R=1/2$, where distinct solutions differ by at least half of the spins, and approximation ratio $a_r = 0.01$. Other classes of instances, as well as the data for distance threshold $R=1/4$, are collected in the Appendix.

For $R=1/2$, all solvers provide a comparable level of sampling fairness, cross-checking that they successfully explore distinct regions of low-energy problem manifold. However, for $R=1/4$, see Appendix~\ref{sec:aux_results}, QA and SBM obtained a significantly larger number of diverse states than the TN approach, which outputs similar numbers as for $R=1/2$ (operating with $M=1024$ in the branch-and-bound search). It may be related to the fact that the D-Wave machine and SBM generally give orders of magnitude more states in their results (equal to the number of samples) than the TN approach, which gets limited by the branch-and-bound search width $M$. As we relax the criterion for diverse states, decreasing $R$, this quantitative advantage naturally translates to more diverse states found.

\section{Contraction stability}
\label{sec:limitations}

\begin{figure}[t!]
\begin{centering}
\includegraphics[width=0.95\columnwidth]{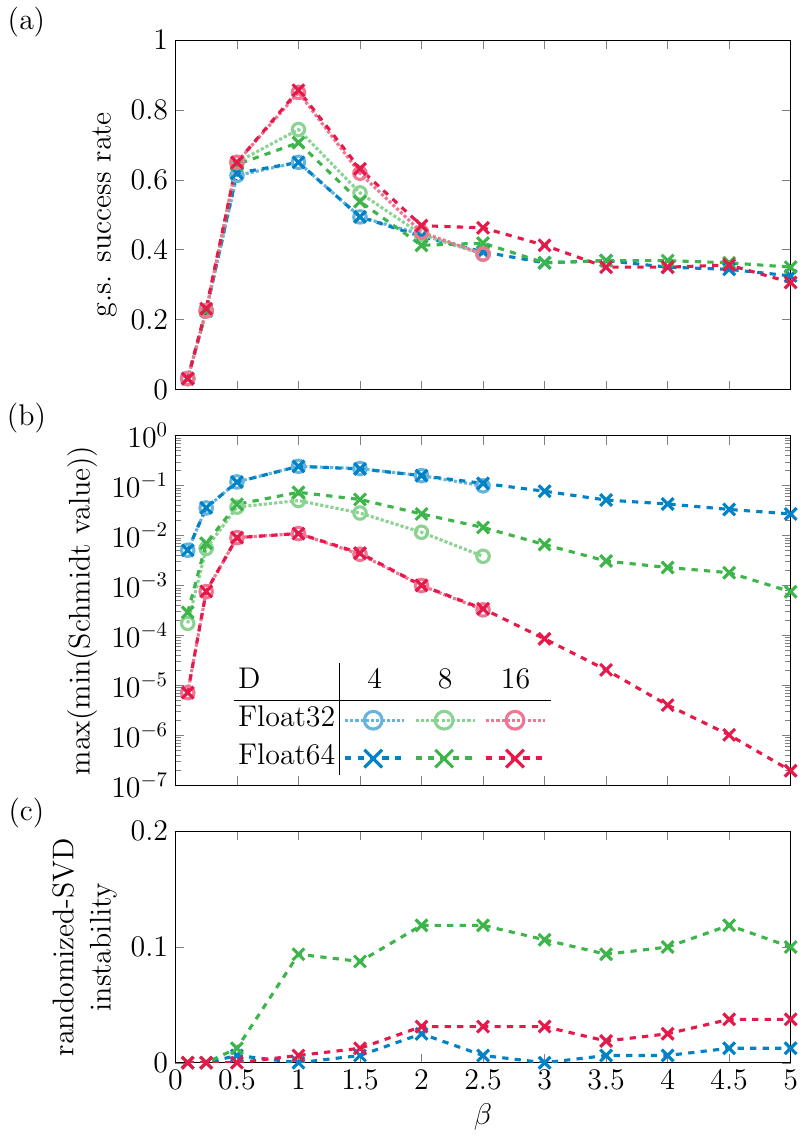}
\par\end{centering}
\caption{Contraction stability. We focus on a minimal Pegasus example with 216 spins.
In (a), we show a percentage of algorithm runs where the ground state has been identified, calculated from 20 problem instances and all 8 rotations (contraction directions) for each instance. We identify the existence of optimal inverse temperature $\beta$, giving a tradeoff between the branch and bound resolution and TN contraction failures. We test the influence of boundary MPS bond dimension $D$ and numerical precision on the latter. In (b), we illustrate the truncation error, estimated from the smallest retained singular values at the worst MPS cut (the median over all instances is plotted).
In (c), we show a percentage of runs where repeated execution gave different outcomes (with respect to finding the ground state). This can be attributed to randomized SVD employed in the algorithm and the inability of variational MPS optimization to correct the resulting fluctuations.
}
\label{fig:stability}
\end{figure}

To diagnose the sources of limitation of the TN approach, we now focus on the smallest problem defined on the Pegasus graph with 216 spins in class I. We test a range of inverse temperatures $\beta$, boundary MPS bond dimensions $D$, and adopted numerical precisions. We employ local dimension reduction to $2^{20}$ to expand the available range of $D$ (we verify that ground states remain supported in reduced spaces). For clarity, in this section, we adopt a metric that tests whether a given algorithm run has found the ground state. The resulting success rate is shown in Fig.~\ref{fig:stability}(a), collecting the data for 20 instances and all possible rotations of each instance.

A general observation is that there is a trade-off between the resolution offered by the branch and bound procedure (where higher $\beta$ is desired) and the stability of TN contraction (where smaller $\beta$ allow for more precise calculation of marginals).  Indeed, for the smallest values of $\beta < 0.25$, the procedure cannot identify the ground state as the branch and bound procedure (here, with $M=1024$) gets stuck in local minima -- independent of contraction parameters.

Those local minima can be avoided by employing larger $\beta$, where, however, the failure of the procedure can be attributed to TN contraction failure. From this perspective, the metric we adopt here provides a direct probe of the stability of the contraction at larger $\beta$. We note that the PEPS tensors we use are built from non-negative elements. In general, this improves contraction stability~\cite{chen_sign_2024} (at the very least, this minimizes numerical-precision errors related to subtracting numbers of similar magnitude). Negative tensor elements, however, do appear in approximate boundary MPS.

Interestingly, in the considered range of parameters, the numerical precision, i.e., a default $64$-bit precision, or a lowered $32$-bit precision used during contraction, does not appreciably affect the results (however, for $32$-bit precision, some underlying numerical operations break for $\beta \ge 3$). This can be understood from Fig~\ref{fig:stability}(b), where we show the truncation accuracy, measured by the smallest retained singular value of boundary MPSs at the worst cut (the cut with the largest such singular value). For $\beta < 3$, 32-bit precision is sufficient to represent those values for $D$ tested. We can also see, following plots at fixed $D$, that with increased $\beta$, the singular values vanish more quickly. This indicates that numerical precision should ultimately become an issue for the largest $\beta$, desired from the perspective of branch and bound.

Finally, we identify that the procedure failure for the biggest $\beta$ is closely tied to randomized truncated SVD~\cite{halko_finding_2009, martinsson_randomized_2020}, which we employ in the calculation of approximate boundary MPSs. Indeed, while the branch and bound procedure, in principle, is deterministic, we see that employing randomized SVD can lead to different outcomes for repeated algorithm runs (with all parameters otherwise fixed). In Fig.~\ref{fig:stability}(c), we plot a percentage of cases where two repeated executions of the algorithm gave different outcomes (in terms of our metric focusing on ground state identification; the instability would be even more pronounced if we compared obtained states). Such an instability is marginal for the smallest values of $\beta$ and the smallest $D=4$. However,  the percentage of unstable runs becomes significant for larger $\beta$ and $D$. 

It indicates that randomized SVD employed in boundary MPS compression results in local optima that the 1-site variational overlap maximalization procedure cannot escape, see Sec.~\ref{sec:zipper}. 
The issue becomes particularly severe when singular values vanish quickly, with a few dominant ones differing by orders of magnitude -- as for $D \ge 8$ at the largest $\beta$ in Fig.~\ref{fig:stability}. This shows that a stable procedure to optimally identify boundary MPS (or, more broadly, contract the network) plays a critical role beyond the bare parameters like the bond dimension D.

The adoption of randomized SVD and variational optimization of boundary MPS limited to 1-site updates is a consequence of the huge dimensions of local spaces and the resulting PEPS bond dimensions. Those appears hard to avoid in the optimization problems we focus on in this article. Any improvement in the randomized SVD performance should translate to improved quality of the results of our TN procedure. This also partially explains the good performance of the algorithm for the Chimera graph in~\cite{rams_approximate_2021}, where standard SVD---and more reliable initialization of boundary MPS optimization---are available, allowing for larger $\beta$. Notwithstanding, we observe that the hardness of contraction grows with lattice connectivity. Typically, the TN approach performs reasonably well for a square lattice, but its performance declines with increased connectivity (including diagonal couplings in a quasi-2D graph). Zephyr seems more complicated than Pegasus, which, in turn, is harder than Chimera or a square lattice.

\section{Discussion}
\label{sec:discussion}

In this work, we implemented a TN-based branch-and-bound search algorithm to identify low-energy states of quasi-2D spin glass problems. We focused on the optimization problems defined on Pegasus and Zephyr graphs, which are currently realized in D-Wave QA. 

We compare the quality of solutions output from our approach with those from the QA and SBM algorithms. The two latter, which can be considered examples of randomized algorithms, operate on approximately two orders of magnitude smaller timescales and, via sampling, provide more dense coverage of the low-energy solutions manifold for considered random problems. A deterministic nature of our branch-and-bound approach likely contributes to the reduced expressiveness of sampling, which provides a possible route for future improvements of that approach.

TN operates best in low-dimensional systems, and in this article, we execute them for problems on the verge of their practical applicability. The limitation to low-dimensional systems is shared by the QA, as we focus here on the optimization problems that utilize the entire interaction graph of the existing hardware. SBM, on the other hand, is not limited by such considerations and can be executed on a fully connected graph.

TN algorithms, including our approach, typically have a sequential nature, making them hard to parallelize, affecting total execution time for a single problem instance. This aspect is shared by a QA, which is producing one classical sample at a time. Still, a single annealing execution time on the order of a millisecond makes them competitive in outputting thousands of samples (though the access time to the online system might add significant overheads). On the other hand, the SBM algorithm, like MCMC approaches, is easily parallelizable, allowing for multiple trajectories to be calculated simultaneously, efficiently utilizing hardware accelerators such as GPU.

Nevertheless, our TN approach is not limited to Ising problems, like the two latter solvers. It can be directly employed for generalized Potts problems (or equivalently Random Markov Fields models) defined on king's graphs (see Fig.~\ref{fig:1}(d)).

We identify the limitations of our approach related to huge bond dimensions appearing in the TN representation of marginals for families of problems we focus on and instabilities in approximate boundary MPS contraction schemes. The performance of our approach might be further improved by introducing additional heuristic TN gauge transformations to stabilize the contraction, which we do not explore in this work.

Our results also illustrate that optimization problems can be employed as stringent tests of the performance of various approximate TN contraction schemes, where the quality of obtained low-energy states directly probes the quality of the contraction, which is a \#P-hard problem in general. 

One emerging area is to build a heterogeneous high-performance quantum-classical computing architecture with adaptive graph partitioning subroutines \cite{Mohseni2024}. This framework could consist of different kind of co-processors that work in concert to accelerate sampling different subgraphs or complementary area of configuration space. This paradigm of hybrid computing could include quantum accelerators (analog quantum annealers or digital QAOA), conventional classical computing accelerators (CPU and GPU), and unconventional classical processors (in-memory \cite{pedretti2023zeroth} and/or probabilistic accelerators \cite{chowdhury2023fullstack}). We  envision that TN approaches can also be incorporated as special deterministic quantum-inspired solvers for sampling low-dimensional subgraphs within such heterogeneous frameworks.

\section*{Data availability}
We make Julia package implementing the algorithm publicly available in~\cite{SGPEPS}. All tested instances and obtained results can be accessed from online repository~\cite{database}.

\acknowledgements{
This project was supported by the National Science Center (NCN), Poland, under projects No.~2020/38/E/ST3/00269 (T.S.), 2022/47/B/ST6/02380 (B.G.), 2020/38/E/ST3/00150 (A.D.), and project 2021/03/Y/ST2/00184 within the QuantERA II Programme that has received funding from the European Union Horizon 2020 research and innovation programme under Grant Agreement No. 101017733 (M.M.R.). This work was in part supported by the Defense Advanced Research Projects Agency (DARPA) under Air Force Research Laboratory (AFRL) contract no. FA8650-23-3-7313 (M.M.).
}

\bibliography{ref.bib}

\begin{thebibliography}{64}%
\makeatletter
\providecommand \@ifxundefined [1]{%
 \@ifx{#1\undefined}
}%
\providecommand \@ifnum [1]{%
 \ifnum #1\expandafter \@firstoftwo
 \else \expandafter \@secondoftwo
 \fi
}%
\providecommand \@ifx [1]{%
 \ifx #1\expandafter \@firstoftwo
 \else \expandafter \@secondoftwo
 \fi
}%
\providecommand \natexlab [1]{#1}%
\providecommand \enquote  [1]{``#1''}%
\providecommand \bibnamefont  [1]{#1}%
\providecommand \bibfnamefont [1]{#1}%
\providecommand \citenamefont [1]{#1}%
\providecommand \href@noop [0]{\@secondoftwo}%
\providecommand \href [0]{\begingroup \@sanitize@url \@href}%
\providecommand \@href[1]{\@@startlink{#1}\@@href}%
\providecommand \@@href[1]{\endgroup#1\@@endlink}%
\providecommand \@sanitize@url [0]{\catcode `\\12\catcode `\$12\catcode
  `\&12\catcode `\#12\catcode `\^12\catcode `\_12\catcode `\%12\relax}%
\providecommand \@@startlink[1]{}%
\providecommand \@@endlink[0]{}%
\providecommand \url  [0]{\begingroup\@sanitize@url \@url }%
\providecommand \@url [1]{\endgroup\@href {#1}{\urlprefix }}%
\providecommand \urlprefix  [0]{URL }%
\providecommand \Eprint [0]{\href }%
\providecommand \doibase [0]{https://doi.org/}%
\providecommand \selectlanguage [0]{\@gobble}%
\providecommand \bibinfo  [0]{\@secondoftwo}%
\providecommand \bibfield  [0]{\@secondoftwo}%
\providecommand \translation [1]{[#1]}%
\providecommand \BibitemOpen [0]{}%
\providecommand \bibitemStop [0]{}%
\providecommand \bibitemNoStop [0]{.\EOS\space}%
\providecommand \EOS [0]{\spacefactor3000\relax}%
\providecommand \BibitemShut  [1]{\csname bibitem#1\endcsname}%
\let\auto@bib@innerbib\@empty
\bibitem [{\citenamefont
  {Barahona}(1982)}]{barahona_on-the-computational_1982}%
  \BibitemOpen
  \bibfield  {author} {\bibinfo {author} {\bibfnamefont {F.}~\bibnamefont
  {Barahona}},\ }\bibfield  {title} {\bibinfo {title} {On the computational
  complexity of {I}sing spin glass models},\ }\href
  {http://stacks.iop.org/0305-4470/15/i=10/a=028} {\bibfield  {journal}
  {\bibinfo  {journal} {J. Phys. A: Math. Gen.}\ }\textbf {\bibinfo {volume}
  {15}},\ \bibinfo {pages} {3241} (\bibinfo {year} {1982})}\BibitemShut
  {NoStop}%
\bibitem [{\citenamefont {Lucas}(2014)}]{lucas_ising_2014}%
  \BibitemOpen
  \bibfield  {author} {\bibinfo {author} {\bibfnamefont {A.}~\bibnamefont
  {Lucas}},\ }\bibfield  {title} {\bibinfo {title} {{Ising} formulations of
  many {NP} problems},\ }\href {https://doi.org/10.3389/fphy.2014.00005}
  {\bibfield  {journal} {\bibinfo  {journal} {Front. Phys.}\ }\textbf {\bibinfo
  {volume} {2}},\ \bibinfo {pages} {5} (\bibinfo {year} {2014})}\BibitemShut
  {NoStop}%
\bibitem [{\citenamefont {De~las Cuevas}\ and\ \citenamefont
  {Cubitt}(2016)}]{de_las_cuevas_simple_2016}%
  \BibitemOpen
  \bibfield  {author} {\bibinfo {author} {\bibfnamefont {G.}~\bibnamefont
  {De~las Cuevas}}\ and\ \bibinfo {author} {\bibfnamefont {T.~S.}\ \bibnamefont
  {Cubitt}},\ }\bibfield  {title} {\bibinfo {title} {Simple universal models
  capture all classical spin physics},\ }\href
  {https://www.sciencemag.org/lookup/doi/10.1126/science.aab3326} {\bibfield
  {journal} {\bibinfo  {journal} {Science}\ }\textbf {\bibinfo {volume}
  {351}},\ \bibinfo {pages} {1180} (\bibinfo {year} {2016})}\BibitemShut
  {NoStop}%
\bibitem [{\citenamefont {Kirkpatrick}\ \emph {et~al.}(1983)\citenamefont
  {Kirkpatrick}, \citenamefont {Gelatt},\ and\ \citenamefont
  {Vecchi}}]{kirkpatrick_optimization_1983}%
  \BibitemOpen
  \bibfield  {author} {\bibinfo {author} {\bibfnamefont {S.}~\bibnamefont
  {Kirkpatrick}}, \bibinfo {author} {\bibfnamefont {C.~D.}\ \bibnamefont
  {Gelatt}},\ and\ \bibinfo {author} {\bibfnamefont {M.~P.}\ \bibnamefont
  {Vecchi}},\ }\bibfield  {title} {\bibinfo {title} {Optimization by simulated
  annealing},\ }\href {https://doi.org/10.1126/science.220.4598.671} {\bibfield
   {journal} {\bibinfo  {journal} {Science}\ }\textbf {\bibinfo {volume}
  {220}},\ \bibinfo {pages} {671} (\bibinfo {year} {1983})}\BibitemShut
  {NoStop}%
\bibitem [{\citenamefont {Metropolis}\ and\ \citenamefont
  {Ulam}(1949)}]{metropolis_monte_1949}%
  \BibitemOpen
  \bibfield  {author} {\bibinfo {author} {\bibfnamefont {N.}~\bibnamefont
  {Metropolis}}\ and\ \bibinfo {author} {\bibfnamefont {S.}~\bibnamefont
  {Ulam}},\ }\bibfield  {title} {\bibinfo {title} {{The Monte Carlo method}},\
  }\href {https://mathscinet.ams.org/mathscinet-getitem?mr=0031341} {\bibfield
  {journal} {\bibinfo  {journal} {J. Amer. Statist. Assoc.}\ }\textbf {\bibinfo
  {volume} {44}},\ \bibinfo {pages} {335–341} (\bibinfo {year}
  {1949})}\BibitemShut {NoStop}%
\bibitem [{\citenamefont {Hastings}(1970)}]{hastings_monte_1970}%
  \BibitemOpen
  \bibfield  {author} {\bibinfo {author} {\bibfnamefont {W.~K.}\ \bibnamefont
  {Hastings}},\ }\bibfield  {title} {\bibinfo {title} {{M}onte {C}arlo sampling
  methods using {M}arkov chains and their applications},\ }\href
  {https://doi.org/10.1093/biomet/57.1.97} {\bibfield  {journal} {\bibinfo
  {journal} {Biometrika}\ }\textbf {\bibinfo {volume} {57}},\ \bibinfo {pages}
  {97} (\bibinfo {year} {1970})}\BibitemShut {NoStop}%
\bibitem [{\citenamefont {Earl}\ and\ \citenamefont
  {Deem}(2005)}]{earl_parallel_tempering_2005}%
  \BibitemOpen
  \bibfield  {author} {\bibinfo {author} {\bibfnamefont {D.~J.}\ \bibnamefont
  {Earl}}\ and\ \bibinfo {author} {\bibfnamefont {M.~W.}\ \bibnamefont
  {Deem}},\ }\bibfield  {title} {\bibinfo {title} {Parallel tempering:
  Theory{,} applications{,} and new perspectives},\ }\href
  {https://doi.org/10.1039/B509983H} {\bibfield  {journal} {\bibinfo  {journal}
  {Phys. Chem. Chem. Phys.}\ }\textbf {\bibinfo {volume} {7}},\ \bibinfo
  {pages} {3910} (\bibinfo {year} {2005})}\BibitemShut {NoStop}%
\bibitem [{\citenamefont {Houdayer}(2001)}]{houdayer_cluster_2001}%
  \BibitemOpen
  \bibfield  {author} {\bibinfo {author} {\bibfnamefont {J.}~\bibnamefont
  {Houdayer}},\ }\bibfield  {title} {\bibinfo {title} {A cluster {M}onte
  {C}arlo algorithm for 2-dimensional spin glasses},\ }\href
  {https://doi.org/10.1007/PL00011151} {\bibfield  {journal} {\bibinfo
  {journal} {Eur. Phys. J. B}\ }\textbf {\bibinfo {volume} {22}},\ \bibinfo
  {pages} {479} (\bibinfo {year} {2001})}\BibitemShut {NoStop}%
\bibitem [{\citenamefont {Zhu}\ \emph {et~al.}(2015)\citenamefont {Zhu},
  \citenamefont {Ochoa},\ and\ \citenamefont
  {Katzgraber}}]{Zhu_efficient_2015}%
  \BibitemOpen
  \bibfield  {author} {\bibinfo {author} {\bibfnamefont {Z.}~\bibnamefont
  {Zhu}}, \bibinfo {author} {\bibfnamefont {A.~J.}\ \bibnamefont {Ochoa}},\
  and\ \bibinfo {author} {\bibfnamefont {H.~G.}\ \bibnamefont {Katzgraber}},\
  }\bibfield  {title} {\bibinfo {title} {Efficient cluster algorithm for spin
  glasses in any space dimension},\ }\href
  {https://doi.org/10.1103/PhysRevLett.115.077201} {\bibfield  {journal}
  {\bibinfo  {journal} {Phys. Rev. Lett.}\ }\textbf {\bibinfo {volume} {115}},\
  \bibinfo {pages} {077201} (\bibinfo {year} {2015})}\BibitemShut {NoStop}%
\bibitem [{\citenamefont {Mohseni}\ \emph {et~al.}(2021)\citenamefont
  {Mohseni}, \citenamefont {Eppens}, \citenamefont {Strumpfer}, \citenamefont
  {Marino}, \citenamefont {Denchev}, \citenamefont {Ho}, \citenamefont
  {Isakov}, \citenamefont {Boixo}, \citenamefont {Ricci-Tersenghi},\ and\
  \citenamefont {Neven}}]{mohseni_nonequilibrium_2021}%
  \BibitemOpen
  \bibfield  {author} {\bibinfo {author} {\bibfnamefont {M.}~\bibnamefont
  {Mohseni}}, \bibinfo {author} {\bibfnamefont {D.}~\bibnamefont {Eppens}},
  \bibinfo {author} {\bibfnamefont {J.}~\bibnamefont {Strumpfer}}, \bibinfo
  {author} {\bibfnamefont {R.}~\bibnamefont {Marino}}, \bibinfo {author}
  {\bibfnamefont {V.}~\bibnamefont {Denchev}}, \bibinfo {author} {\bibfnamefont
  {A.~K.}\ \bibnamefont {Ho}}, \bibinfo {author} {\bibfnamefont {S.~V.}\
  \bibnamefont {Isakov}}, \bibinfo {author} {\bibfnamefont {S.}~\bibnamefont
  {Boixo}}, \bibinfo {author} {\bibfnamefont {F.}~\bibnamefont
  {Ricci-Tersenghi}},\ and\ \bibinfo {author} {\bibfnamefont {H.}~\bibnamefont
  {Neven}},\ }\bibfield  {title} {\bibinfo {title} {Nonequilibrium {M}onte
  {C}arlo for unfreezing variables in hard combinatorial optimization},\ }\href
  {https://doi.org/10.48550/arXiv.2111.13628} {\bibfield  {journal} {\bibinfo
  {journal} {arXiv:2111.13628}\ } (\bibinfo {year} {2021})}\BibitemShut
  {NoStop}%
\bibitem [{\citenamefont {Selby}(2014)}]{selby_efficient_2014}%
  \BibitemOpen
  \bibfield  {author} {\bibinfo {author} {\bibfnamefont {A.}~\bibnamefont
  {Selby}},\ }\bibfield  {title} {\bibinfo {title} {Efficient subgraph-based
  sampling of {I}sing-type models with frustration},\ }\href
  {https://doi.org/10.48550/arXiv.1409.3934} {\bibfield  {journal} {\bibinfo
  {journal} {arXiv:1409.3934}\ } (\bibinfo {year} {2014})}\BibitemShut
  {NoStop}%
\bibitem [{\citenamefont {Isakov}\ \emph {et~al.}(2015)\citenamefont {Isakov},
  \citenamefont {Zintchenko}, \citenamefont {Rønnow},\ and\ \citenamefont
  {Troyer}}]{isakov_optimised_2015}%
  \BibitemOpen
  \bibfield  {author} {\bibinfo {author} {\bibfnamefont {S.}~\bibnamefont
  {Isakov}}, \bibinfo {author} {\bibfnamefont {I.}~\bibnamefont {Zintchenko}},
  \bibinfo {author} {\bibfnamefont {T.}~\bibnamefont {Rønnow}},\ and\ \bibinfo
  {author} {\bibfnamefont {M.}~\bibnamefont {Troyer}},\ }\bibfield  {title}
  {\bibinfo {title} {Optimised simulated annealing for {I}sing spin glasses},\
  }\href {https://doi.org/https://doi.org/10.1016/j.cpc.2015.02.015} {\bibfield
   {journal} {\bibinfo  {journal} {Comput. Phys. Commun.}\ }\textbf {\bibinfo
  {volume} {192}},\ \bibinfo {pages} {265} (\bibinfo {year}
  {2015})}\BibitemShut {NoStop}%
\bibitem [{\citenamefont {Kadowaki}\ and\ \citenamefont
  {Nishimori}(1998)}]{Kadowaki_quantum_1998}%
  \BibitemOpen
  \bibfield  {author} {\bibinfo {author} {\bibfnamefont {T.}~\bibnamefont
  {Kadowaki}}\ and\ \bibinfo {author} {\bibfnamefont {H.}~\bibnamefont
  {Nishimori}},\ }\bibfield  {title} {\bibinfo {title} {Quantum annealing in
  the transverse {I}sing model},\ }\href
  {https://doi.org/10.1103/PhysRevE.58.5355} {\bibfield  {journal} {\bibinfo
  {journal} {Phys. Rev. E}\ }\textbf {\bibinfo {volume} {58}},\ \bibinfo
  {pages} {5355} (\bibinfo {year} {1998})}\BibitemShut {NoStop}%
\bibitem [{\citenamefont {Santoro}\ \emph {et~al.}(2002)\citenamefont
  {Santoro}, \citenamefont {Martoňák}, \citenamefont {Tosatti},\ and\
  \citenamefont {Car}}]{Santoro_theory_2002}%
  \BibitemOpen
  \bibfield  {author} {\bibinfo {author} {\bibfnamefont {G.~E.}\ \bibnamefont
  {Santoro}}, \bibinfo {author} {\bibfnamefont {R.}~\bibnamefont {Martoňák}},
  \bibinfo {author} {\bibfnamefont {E.}~\bibnamefont {Tosatti}},\ and\ \bibinfo
  {author} {\bibfnamefont {R.}~\bibnamefont {Car}},\ }\bibfield  {title}
  {\bibinfo {title} {Theory of quantum annealing of an {I}sing spin glass},\
  }\href {https://doi.org/10.1126/science.1068774} {\bibfield  {journal}
  {\bibinfo  {journal} {Science}\ }\textbf {\bibinfo {volume} {295}},\ \bibinfo
  {pages} {2427} (\bibinfo {year} {2002})}\BibitemShut {NoStop}%
\bibitem [{\citenamefont {King}\ \emph {et~al.}(2023)\citenamefont {King},
  \citenamefont {Raymond}, \citenamefont {Lanting}, \citenamefont {Harris},
  \citenamefont {Zucca}, \citenamefont {Altomare}, \citenamefont {Berkley},
  \citenamefont {Boothby}, \citenamefont {Ejtemaee}, \citenamefont {Enderud}
  \emph {et~al.}}]{King_quantum_2023}%
  \BibitemOpen
  \bibfield  {author} {\bibinfo {author} {\bibfnamefont {A.~D.}\ \bibnamefont
  {King}}, \bibinfo {author} {\bibfnamefont {J.}~\bibnamefont {Raymond}},
  \bibinfo {author} {\bibfnamefont {T.}~\bibnamefont {Lanting}}, \bibinfo
  {author} {\bibfnamefont {R.}~\bibnamefont {Harris}}, \bibinfo {author}
  {\bibfnamefont {A.}~\bibnamefont {Zucca}}, \bibinfo {author} {\bibfnamefont
  {F.}~\bibnamefont {Altomare}}, \bibinfo {author} {\bibfnamefont {A.~J.}\
  \bibnamefont {Berkley}}, \bibinfo {author} {\bibfnamefont {K.}~\bibnamefont
  {Boothby}}, \bibinfo {author} {\bibfnamefont {S.}~\bibnamefont {Ejtemaee}},
  \bibinfo {author} {\bibfnamefont {C.}~\bibnamefont {Enderud}}, \emph
  {et~al.},\ }\bibfield  {title} {\bibinfo {title} {Quantum critical dynamics
  in a 5,000-qubit programmable spin glass},\ }\href
  {https://doi.org/10.1038/s41586-023-05867-2} {\bibfield  {journal} {\bibinfo
  {journal} {Nature}\ }\textbf {\bibinfo {volume} {617}},\ \bibinfo {pages}
  {61} (\bibinfo {year} {2023})}\BibitemShut {NoStop}%
\bibitem [{\citenamefont {Bernaschi}\ \emph {et~al.}(2024)\citenamefont
  {Bernaschi}, \citenamefont {González-Adalid~Pemartín}, \citenamefont
  {Martín-Mayor},\ and\ \citenamefont {Parisi}}]{bernaschi_quantum_2024}%
  \BibitemOpen
  \bibfield  {author} {\bibinfo {author} {\bibfnamefont {M.}~\bibnamefont
  {Bernaschi}}, \bibinfo {author} {\bibfnamefont {I.}~\bibnamefont
  {González-Adalid~Pemartín}}, \bibinfo {author} {\bibfnamefont
  {V.}~\bibnamefont {Martín-Mayor}},\ and\ \bibinfo {author} {\bibfnamefont
  {G.}~\bibnamefont {Parisi}},\ }\bibfield  {title} {\bibinfo {title} {The
  quantum transition of the two-dimensional {I}sing spin glass},\ }\href
  {https://doi.org/10.1038/s41586-024-07647-y} {\bibfield  {journal} {\bibinfo
  {journal} {Nature}\ }\textbf {\bibinfo {volume} {631}},\ \bibinfo {pages}
  {749} (\bibinfo {year} {2024})}\BibitemShut {NoStop}%
\bibitem [{\citenamefont {Munoz-Bauza}\ and\ \citenamefont
  {Lidar}(2025)}]{bauza_scaling_2024}%
  \BibitemOpen
  \bibfield  {author} {\bibinfo {author} {\bibfnamefont {H.}~\bibnamefont
  {Munoz-Bauza}}\ and\ \bibinfo {author} {\bibfnamefont {D.}~\bibnamefont
  {Lidar}},\ }\bibfield  {title} {\bibinfo {title} {Scaling advantage in
  approximate optimization with quantum annealing},\ }\href
  {https://doi.org/10.1103/PhysRevLett.134.160601} {\bibfield  {journal}
  {\bibinfo  {journal} {Phys. Rev. Lett.}\ }\textbf {\bibinfo {volume} {134}},\
  \bibinfo {pages} {160601} (\bibinfo {year} {2025})}\BibitemShut {NoStop}%
\bibitem [{\citenamefont {King}\ \emph {et~al.}(2022)\citenamefont {King},
  \citenamefont {Suzuki}, \citenamefont {Raymond}, \citenamefont {Zucca},
  \citenamefont {Lanting}, \citenamefont {Altomare}, \citenamefont {Berkley},
  \citenamefont {Ejtemaee}, \citenamefont {Hoskinson}, \citenamefont {Huang}
  \emph {et~al.}}]{king_coherent_2022}%
  \BibitemOpen
  \bibfield  {author} {\bibinfo {author} {\bibfnamefont {A.~D.}\ \bibnamefont
  {King}}, \bibinfo {author} {\bibfnamefont {S.}~\bibnamefont {Suzuki}},
  \bibinfo {author} {\bibfnamefont {J.}~\bibnamefont {Raymond}}, \bibinfo
  {author} {\bibfnamefont {A.}~\bibnamefont {Zucca}}, \bibinfo {author}
  {\bibfnamefont {T.}~\bibnamefont {Lanting}}, \bibinfo {author} {\bibfnamefont
  {F.}~\bibnamefont {Altomare}}, \bibinfo {author} {\bibfnamefont {A.~J.}\
  \bibnamefont {Berkley}}, \bibinfo {author} {\bibfnamefont {S.}~\bibnamefont
  {Ejtemaee}}, \bibinfo {author} {\bibfnamefont {E.}~\bibnamefont {Hoskinson}},
  \bibinfo {author} {\bibfnamefont {S.}~\bibnamefont {Huang}}, \emph {et~al.},\
  }\bibfield  {title} {\bibinfo {title} {Coherent quantum annealing in a
  programmable 2000-qubit {Ising} chain},\ }\href
  {https://doi.org/10.1038/s41567-022-01741-6} {\bibfield  {journal} {\bibinfo
  {journal} {Nat. Phys.}\ }\textbf {\bibinfo {volume} {18}},\ \bibinfo {pages}
  {1324} (\bibinfo {year} {2022})}\BibitemShut {NoStop}%
\bibitem [{\citenamefont {King}\ \emph {et~al.}(2025)\citenamefont {King},
  \citenamefont {Nocera}, \citenamefont {Rams}, \citenamefont {Dziarmaga},
  \citenamefont {Wiersema}, \citenamefont {Bernoudy}, \citenamefont {Raymond},
  \citenamefont {Kaushal}, \citenamefont {Heinsdorf}, \citenamefont {Harris}
  \emph {et~al.}}]{king_computational_2024}%
  \BibitemOpen
  \bibfield  {author} {\bibinfo {author} {\bibfnamefont {A.~D.}\ \bibnamefont
  {King}}, \bibinfo {author} {\bibfnamefont {A.}~\bibnamefont {Nocera}},
  \bibinfo {author} {\bibfnamefont {M.~M.}\ \bibnamefont {Rams}}, \bibinfo
  {author} {\bibfnamefont {J.}~\bibnamefont {Dziarmaga}}, \bibinfo {author}
  {\bibfnamefont {R.}~\bibnamefont {Wiersema}}, \bibinfo {author}
  {\bibfnamefont {W.}~\bibnamefont {Bernoudy}}, \bibinfo {author}
  {\bibfnamefont {J.}~\bibnamefont {Raymond}}, \bibinfo {author} {\bibfnamefont
  {N.}~\bibnamefont {Kaushal}}, \bibinfo {author} {\bibfnamefont
  {N.}~\bibnamefont {Heinsdorf}}, \bibinfo {author} {\bibfnamefont
  {R.}~\bibnamefont {Harris}}, \emph {et~al.},\ }\bibfield  {title} {\bibinfo
  {title} {Beyond-classical computation in quantum simulation},\ }\href
  {https://doi.org/https://doi.org/10.1126/science.ado6285} {\bibfield
  {journal} {\bibinfo  {journal} {Science}\ }\textbf {\bibinfo {volume}
  {388}},\ \bibinfo {pages} {199} (\bibinfo {year} {2025})}\BibitemShut
  {NoStop}%
\bibitem [{\citenamefont {Bando}\ \emph {et~al.}(2020)\citenamefont {Bando},
  \citenamefont {Susa}, \citenamefont {Oshiyama}, \citenamefont {Shibata},
  \citenamefont {Ohzeki}, \citenamefont {G\'omez-Ruiz}, \citenamefont {Lidar},
  \citenamefont {Suzuki}, \citenamefont {del Campo},\ and\ \citenamefont
  {Nishimori}}]{bando_probing_2020}%
  \BibitemOpen
  \bibfield  {author} {\bibinfo {author} {\bibfnamefont {Y.}~\bibnamefont
  {Bando}}, \bibinfo {author} {\bibfnamefont {Y.}~\bibnamefont {Susa}},
  \bibinfo {author} {\bibfnamefont {H.}~\bibnamefont {Oshiyama}}, \bibinfo
  {author} {\bibfnamefont {N.}~\bibnamefont {Shibata}}, \bibinfo {author}
  {\bibfnamefont {M.}~\bibnamefont {Ohzeki}}, \bibinfo {author} {\bibfnamefont
  {F.~J.}\ \bibnamefont {G\'omez-Ruiz}}, \bibinfo {author} {\bibfnamefont
  {D.~A.}\ \bibnamefont {Lidar}}, \bibinfo {author} {\bibfnamefont
  {S.}~\bibnamefont {Suzuki}}, \bibinfo {author} {\bibfnamefont
  {A.}~\bibnamefont {del Campo}},\ and\ \bibinfo {author} {\bibfnamefont
  {H.}~\bibnamefont {Nishimori}},\ }\bibfield  {title} {\bibinfo {title}
  {Probing the universality of topological defect formation in a quantum
  annealer: {K}ibble-{Z}urek mechanism and beyond},\ }\href
  {https://doi.org/10.1103/PhysRevResearch.2.033369} {\bibfield  {journal}
  {\bibinfo  {journal} {Phys. Rev. Res.}\ }\textbf {\bibinfo {volume} {2}},\
  \bibinfo {pages} {033369} (\bibinfo {year} {2020})}\BibitemShut {NoStop}%
\bibitem [{\citenamefont {Kashimata}\ \emph {et~al.}(2024)\citenamefont
  {Kashimata}, \citenamefont {Yamasaki}, \citenamefont {Hidaka},\ and\
  \citenamefont {Tatsumura}}]{kashimata_efficient_2024}%
  \BibitemOpen
  \bibfield  {author} {\bibinfo {author} {\bibfnamefont {T.}~\bibnamefont
  {Kashimata}}, \bibinfo {author} {\bibfnamefont {M.}~\bibnamefont {Yamasaki}},
  \bibinfo {author} {\bibfnamefont {R.}~\bibnamefont {Hidaka}},\ and\ \bibinfo
  {author} {\bibfnamefont {K.}~\bibnamefont {Tatsumura}},\ }\bibfield  {title}
  {\bibinfo {title} {Efficient and scalable architecture for multiple-chip
  implementation of simulated bifurcation machines},\ }\href
  {https://doi.org/10.1109/access.2024.3374089} {\bibfield  {journal} {\bibinfo
   {journal} {IEEE Access}\ }\textbf {\bibinfo {volume} {12}},\ \bibinfo
  {pages} {36606–36621} (\bibinfo {year} {2024})}\BibitemShut {NoStop}%
\bibitem [{\citenamefont {Goto}\ \emph {et~al.}(2019)\citenamefont {Goto},
  \citenamefont {Tatsumura},\ and\ \citenamefont
  {Dixon}}]{goto_combinatorial_2019}%
  \BibitemOpen
  \bibfield  {author} {\bibinfo {author} {\bibfnamefont {H.}~\bibnamefont
  {Goto}}, \bibinfo {author} {\bibfnamefont {K.}~\bibnamefont {Tatsumura}},\
  and\ \bibinfo {author} {\bibfnamefont {A.~R.}\ \bibnamefont {Dixon}},\
  }\bibfield  {title} {\bibinfo {title} {Combinatorial optimization by
  simulating adiabatic bifurcations in nonlinear {{Hamiltonian}} systems},\
  }\href {https://doi.org/10.1126/sciadv.aav2372} {\bibfield  {journal}
  {\bibinfo  {journal} {Sci. Adv.}\ }\textbf {\bibinfo {volume} {5}},\ \bibinfo
  {pages} {eaav2372} (\bibinfo {year} {2019})}\BibitemShut {NoStop}%
\bibitem [{\citenamefont {Goto}\ \emph {et~al.}(2021)\citenamefont {Goto},
  \citenamefont {Endo}, \citenamefont {Suzuki}, \citenamefont {Sakai},
  \citenamefont {Kanao}, \citenamefont {Hamakawa}, \citenamefont {Hidaka},
  \citenamefont {Yamasaki},\ and\ \citenamefont
  {Tatsumura}}]{goto_highperformance_2021}%
  \BibitemOpen
  \bibfield  {author} {\bibinfo {author} {\bibfnamefont {H.}~\bibnamefont
  {Goto}}, \bibinfo {author} {\bibfnamefont {K.}~\bibnamefont {Endo}}, \bibinfo
  {author} {\bibfnamefont {M.}~\bibnamefont {Suzuki}}, \bibinfo {author}
  {\bibfnamefont {Y.}~\bibnamefont {Sakai}}, \bibinfo {author} {\bibfnamefont
  {T.}~\bibnamefont {Kanao}}, \bibinfo {author} {\bibfnamefont
  {Y.}~\bibnamefont {Hamakawa}}, \bibinfo {author} {\bibfnamefont
  {R.}~\bibnamefont {Hidaka}}, \bibinfo {author} {\bibfnamefont
  {M.}~\bibnamefont {Yamasaki}},\ and\ \bibinfo {author} {\bibfnamefont
  {K.}~\bibnamefont {Tatsumura}},\ }\bibfield  {title} {\bibinfo {title}
  {High-performance combinatorial optimization based on classical mechanics},\
  }\href {https://doi.org/10.1126/sciadv.abe7953} {\bibfield  {journal}
  {\bibinfo  {journal} {Sci. Adv.}\ }\textbf {\bibinfo {volume} {7}},\ \bibinfo
  {pages} {eabe7953} (\bibinfo {year} {2021})}\BibitemShut {NoStop}%
\bibitem [{\citenamefont {Verstraete}\ \emph {et~al.}(2008)\citenamefont
  {Verstraete}, \citenamefont {Murg},\ and\ \citenamefont
  {Cirac}}]{verstraete_matrix_2008}%
  \BibitemOpen
  \bibfield  {author} {\bibinfo {author} {\bibfnamefont {F.}~\bibnamefont
  {Verstraete}}, \bibinfo {author} {\bibfnamefont {V.}~\bibnamefont {Murg}},\
  and\ \bibinfo {author} {\bibfnamefont {J.~I.}\ \bibnamefont {Cirac}},\
  }\bibfield  {title} {\bibinfo {title} {Matrix {Product} {States}, {Projected}
  {Entangled} {Pair} {States}, and variational renormalization group methods
  for quantum spin systems},\ }\href
  {http://www.tandfonline.com/doi/abs/10.1080/14789940801912366} {\bibfield
  {journal} {\bibinfo  {journal} {Adv. Phys.}\ }\textbf {\bibinfo {volume}
  {57}},\ \bibinfo {pages} {143} (\bibinfo {year} {2008})}\BibitemShut
  {NoStop}%
\bibitem [{\citenamefont {Or{\'u}s}(2014)}]{orus_practical_2014}%
  \BibitemOpen
  \bibfield  {author} {\bibinfo {author} {\bibfnamefont {R.}~\bibnamefont
  {Or{\'u}s}},\ }\bibfield  {title} {\bibinfo {title} {A practical introduction
  to tensor networks: Matrix product states and projected entangled pair
  states},\ }\href {https://doi.org/https://doi.org/10.1016/j.aop.2014.06.013}
  {\bibfield  {journal} {\bibinfo  {journal} {Ann. Phys.}\ }\textbf {\bibinfo
  {volume} {349}},\ \bibinfo {pages} {117 } (\bibinfo {year}
  {2014})}\BibitemShut {NoStop}%
\bibitem [{\citenamefont {{Biamonte}}\ and\ \citenamefont
  {{Bergholm}}(2017)}]{biamonte_tensor_2017}%
  \BibitemOpen
  \bibfield  {author} {\bibinfo {author} {\bibfnamefont {J.}~\bibnamefont
  {{Biamonte}}}\ and\ \bibinfo {author} {\bibfnamefont {V.}~\bibnamefont
  {{Bergholm}}},\ }\bibfield  {title} {\bibinfo {title} {Tensor networks in a
  nutshell},\ }\href {https://doi.org/10.48550/arXiv.1708.00006} {\bibfield
  {journal} {\bibinfo  {journal} {arXiv:1708.00006}\ } (\bibinfo {year}
  {2017})}\BibitemShut {NoStop}%
\bibitem [{\citenamefont {Ran}\ \emph {et~al.}(2020)\citenamefont {Ran},
  \citenamefont {Tirrito}, \citenamefont {Peng}, \citenamefont {Chen},
  \citenamefont {Su},\ and\ \citenamefont {Lewenstein}}]{ran_tensor_2020}%
  \BibitemOpen
  \bibfield  {author} {\bibinfo {author} {\bibfnamefont {S.-J.}\ \bibnamefont
  {Ran}}, \bibinfo {author} {\bibfnamefont {E.}~\bibnamefont {Tirrito}},
  \bibinfo {author} {\bibfnamefont {C.}~\bibnamefont {Peng}}, \bibinfo {author}
  {\bibfnamefont {X.}~\bibnamefont {Chen}}, \bibinfo {author} {\bibfnamefont
  {G.}~\bibnamefont {Su}},\ and\ \bibinfo {author} {\bibfnamefont
  {M.}~\bibnamefont {Lewenstein}},\ }\href
  {https://link.springer.com/book/10.1007%2F978-3-030-34489-4} {\emph {\bibinfo
  {title} {Tensor Network Contractions}}}\ (\bibinfo  {publisher} {Springer},\
  \bibinfo {year} {2020})\BibitemShut {NoStop}%
\bibitem [{\citenamefont {Bañuls}(2023)}]{banuls_tensor_2023}%
  \BibitemOpen
  \bibfield  {author} {\bibinfo {author} {\bibfnamefont {M.~C.}\ \bibnamefont
  {Bañuls}},\ }\bibfield  {title} {\bibinfo {title} {Tensor network
  algorithms: A route map},\ }\href
  {https://doi.org/https://doi.org/10.1146/annurev-conmatphys-040721-022705}
  {\bibfield  {journal} {\bibinfo  {journal} {Annu. Rev. Condens. Matter
  Phys.}\ }\textbf {\bibinfo {volume} {14}},\ \bibinfo {pages} {173} (\bibinfo
  {year} {2023})}\BibitemShut {NoStop}%
\bibitem [{\citenamefont {Nishino}\ \emph {et~al.}(2001)\citenamefont
  {Nishino}, \citenamefont {Hieida}, \citenamefont {Okunishi}, \citenamefont
  {Maeshima}, \citenamefont {Akutsu},\ and\ \citenamefont
  {Gendiar}}]{nishino_two-dimensional_2001}%
  \BibitemOpen
  \bibfield  {author} {\bibinfo {author} {\bibfnamefont {T.}~\bibnamefont
  {Nishino}}, \bibinfo {author} {\bibfnamefont {Y.}~\bibnamefont {Hieida}},
  \bibinfo {author} {\bibfnamefont {K.}~\bibnamefont {Okunishi}}, \bibinfo
  {author} {\bibfnamefont {N.}~\bibnamefont {Maeshima}}, \bibinfo {author}
  {\bibfnamefont {Y.}~\bibnamefont {Akutsu}},\ and\ \bibinfo {author}
  {\bibfnamefont {A.}~\bibnamefont {Gendiar}},\ }\bibfield  {title} {\bibinfo
  {title} {Two-dimensional tensor product variational formulation},\ }\href
  {https://doi.org/10.1143/PTP.105.409} {\bibfield  {journal} {\bibinfo
  {journal} {Progr. Theor. Phys.}\ }\textbf {\bibinfo {volume} {105}},\
  \bibinfo {pages} {409} (\bibinfo {year} {2001})}\BibitemShut {NoStop}%
\bibitem [{\citenamefont {Schuch}\ \emph {et~al.}(2007)\citenamefont {Schuch},
  \citenamefont {Wolf}, \citenamefont {Verstraete},\ and\ \citenamefont
  {Cirac}}]{schuch_computational_2007}%
  \BibitemOpen
  \bibfield  {author} {\bibinfo {author} {\bibfnamefont {N.}~\bibnamefont
  {Schuch}}, \bibinfo {author} {\bibfnamefont {M.~M.}\ \bibnamefont {Wolf}},
  \bibinfo {author} {\bibfnamefont {F.}~\bibnamefont {Verstraete}},\ and\
  \bibinfo {author} {\bibfnamefont {J.~I.}\ \bibnamefont {Cirac}},\ }\bibfield
  {title} {\bibinfo {title} {Computational complexity of projected entangled
  pair states},\ }\href {https://doi.org/10.1103/PhysRevLett.98.140506}
  {\bibfield  {journal} {\bibinfo  {journal} {Phys. Rev. Lett.}\ }\textbf
  {\bibinfo {volume} {98}},\ \bibinfo {pages} {140506} (\bibinfo {year}
  {2007})}\BibitemShut {NoStop}%
\bibitem [{\citenamefont {Rams}\ \emph {et~al.}(2021)\citenamefont {Rams},
  \citenamefont {Mohseni}, \citenamefont {Eppens}, \citenamefont
  {Ja\l{}owiecki},\ and\ \citenamefont {Gardas}}]{rams_approximate_2021}%
  \BibitemOpen
  \bibfield  {author} {\bibinfo {author} {\bibfnamefont {M.~M.}\ \bibnamefont
  {Rams}}, \bibinfo {author} {\bibfnamefont {M.}~\bibnamefont {Mohseni}},
  \bibinfo {author} {\bibfnamefont {D.}~\bibnamefont {Eppens}}, \bibinfo
  {author} {\bibfnamefont {K.}~\bibnamefont {Ja\l{}owiecki}},\ and\ \bibinfo
  {author} {\bibfnamefont {B.}~\bibnamefont {Gardas}},\ }\bibfield  {title}
  {\bibinfo {title} {Approximate optimization, sampling, and spin-glass droplet
  discovery with tensor networks},\ }\href
  {https://doi.org/10.1103/PhysRevE.104.025308} {\bibfield  {journal} {\bibinfo
   {journal} {Phys. Rev. E}\ }\textbf {\bibinfo {volume} {104}},\ \bibinfo
  {pages} {025308} (\bibinfo {year} {2021})}\BibitemShut {NoStop}%
\bibitem [{\citenamefont {Ueda}\ \emph {et~al.}(2005)\citenamefont {Ueda},
  \citenamefont {Otani}, \citenamefont {Nishio}, \citenamefont {Gendiar},\ and\
  \citenamefont {Nishino}}]{ueda_snapshot_2005}%
  \BibitemOpen
  \bibfield  {author} {\bibinfo {author} {\bibfnamefont {K.}~\bibnamefont
  {Ueda}}, \bibinfo {author} {\bibfnamefont {R.}~\bibnamefont {Otani}},
  \bibinfo {author} {\bibfnamefont {Y.}~\bibnamefont {Nishio}}, \bibinfo
  {author} {\bibfnamefont {A.}~\bibnamefont {Gendiar}},\ and\ \bibinfo {author}
  {\bibfnamefont {T.}~\bibnamefont {Nishino}},\ }\bibfield  {title} {\bibinfo
  {title} {Snapshot observation for {2D} classical lattice models by corner
  transfer matrix renormalization group},\ }\href
  {https://doi.org/10.1143/JPSJS.74S.111} {\bibfield  {journal} {\bibinfo
  {journal} {J. Phys. Soc. Jap.}\ }\textbf {\bibinfo {volume} {74}},\ \bibinfo
  {pages} {111} (\bibinfo {year} {2005})}\BibitemShut {NoStop}%
\bibitem [{\citenamefont {Frias-Perez}\ \emph {et~al.}(2023)\citenamefont
  {Frias-Perez}, \citenamefont {Marien}, \citenamefont {Garcia}, \citenamefont
  {Banuls},\ and\ \citenamefont {Iblisdir}}]{FriasPerez_collective_2023}%
  \BibitemOpen
  \bibfield  {author} {\bibinfo {author} {\bibfnamefont {M.}~\bibnamefont
  {Frias-Perez}}, \bibinfo {author} {\bibfnamefont {M.}~\bibnamefont {Marien}},
  \bibinfo {author} {\bibfnamefont {D.~P.}\ \bibnamefont {Garcia}}, \bibinfo
  {author} {\bibfnamefont {M.~C.}\ \bibnamefont {Banuls}},\ and\ \bibinfo
  {author} {\bibfnamefont {S.}~\bibnamefont {Iblisdir}},\ }\bibfield  {title}
  {\bibinfo {title} {{Collective Monte Carlo updates through tensor network
  renormalization}},\ }\href {https://doi.org/10.21468/SciPostPhys.14.5.123}
  {\bibfield  {journal} {\bibinfo  {journal} {SciPost Phys.}\ }\textbf
  {\bibinfo {volume} {14}},\ \bibinfo {pages} {123} (\bibinfo {year}
  {2023})}\BibitemShut {NoStop}%
\bibitem [{\citenamefont {Gangat}\ and\ \citenamefont
  {Gray}(2024)}]{gangat_hyperoptimize_2024}%
  \BibitemOpen
  \bibfield  {author} {\bibinfo {author} {\bibfnamefont {A.~A.}\ \bibnamefont
  {Gangat}}\ and\ \bibinfo {author} {\bibfnamefont {J.}~\bibnamefont {Gray}},\
  }\bibfield  {title} {\bibinfo {title} {Hyperoptimized approximate contraction
  of tensor networks for rugged-energy-landscape spin glasses on periodic
  square and cubic lattices},\ }\href
  {https://doi.org/10.1103/PhysRevE.110.065306} {\bibfield  {journal} {\bibinfo
   {journal} {Phys. Rev. E}\ }\textbf {\bibinfo {volume} {110}},\ \bibinfo
  {pages} {065306} (\bibinfo {year} {2024})}\BibitemShut {NoStop}%
\bibitem [{\citenamefont {Gray}\ and\ \citenamefont
  {Chan}(2024)}]{gray_hyperoptimized_2024}%
  \BibitemOpen
  \bibfield  {author} {\bibinfo {author} {\bibfnamefont {J.}~\bibnamefont
  {Gray}}\ and\ \bibinfo {author} {\bibfnamefont {G.~K.-L.}\ \bibnamefont
  {Chan}},\ }\bibfield  {title} {\bibinfo {title} {Hyperoptimized approximate
  contraction of tensor networks with arbitrary geometry},\ }\href
  {https://doi.org/10.1103/PhysRevX.14.011009} {\bibfield  {journal} {\bibinfo
  {journal} {Phys. Rev. X}\ }\textbf {\bibinfo {volume} {14}},\ \bibinfo
  {pages} {011009} (\bibinfo {year} {2024})}\BibitemShut {NoStop}%
\bibitem [{\citenamefont {Liu}\ \emph {et~al.}(2021)\citenamefont {Liu},
  \citenamefont {Wang},\ and\ \citenamefont {Zhang}}]{liu_tropical_2021}%
  \BibitemOpen
  \bibfield  {author} {\bibinfo {author} {\bibfnamefont {J.-G.}\ \bibnamefont
  {Liu}}, \bibinfo {author} {\bibfnamefont {L.}~\bibnamefont {Wang}},\ and\
  \bibinfo {author} {\bibfnamefont {P.}~\bibnamefont {Zhang}},\ }\bibfield
  {title} {\bibinfo {title} {Tropical tensor network for ground states of spin
  glasses},\ }\href {https://doi.org/10.1103/PhysRevLett.126.090506} {\bibfield
   {journal} {\bibinfo  {journal} {Phys. Rev. Lett.}\ }\textbf {\bibinfo
  {volume} {126}},\ \bibinfo {pages} {090506} (\bibinfo {year}
  {2021})}\BibitemShut {NoStop}%
\bibitem [{\citenamefont {Mezard}\ and\ \citenamefont
  {Montanari}(2009)}]{mezard_information_2009}%
  \BibitemOpen
  \bibfield  {author} {\bibinfo {author} {\bibfnamefont {M.}~\bibnamefont
  {Mezard}}\ and\ \bibinfo {author} {\bibfnamefont {A.}~\bibnamefont
  {Montanari}},\ }\href@noop {} {\emph {\bibinfo {title} {Information, physics,
  and computation}}},\ Oxford graduate texts\ (\bibinfo  {publisher} {Oxford
  University Press},\ \bibinfo {address} {Oxford ; New York},\ \bibinfo {year}
  {2009})\BibitemShut {NoStop}%
\bibitem [{\citenamefont {Dattani}\ \emph {et~al.}(2019)\citenamefont
  {Dattani}, \citenamefont {Szalay},\ and\ \citenamefont
  {Chancellor}}]{dattani_pegasus_2019}%
  \BibitemOpen
  \bibfield  {author} {\bibinfo {author} {\bibfnamefont {N.}~\bibnamefont
  {Dattani}}, \bibinfo {author} {\bibfnamefont {S.}~\bibnamefont {Szalay}},\
  and\ \bibinfo {author} {\bibfnamefont {N.}~\bibnamefont {Chancellor}},\
  }\bibfield  {title} {\bibinfo {title} {Pegasus: The second connectivity graph
  for large-scale quantum annealing hardware},\ }\href
  {https://doi.org/10.48550/arXiv.1901.07636} {\bibfield  {journal} {\bibinfo
  {journal} {arXiv:1901.07636}\ } (\bibinfo {year} {2019})}\BibitemShut
  {NoStop}%
\bibitem [{\citenamefont {Boothby}\ \emph {et~al.}(2020)\citenamefont
  {Boothby}, \citenamefont {Bunyk}, \citenamefont {Raymond},\ and\
  \citenamefont {Roy}}]{boothby_next-generation_2020}%
  \BibitemOpen
  \bibfield  {author} {\bibinfo {author} {\bibfnamefont {K.}~\bibnamefont
  {Boothby}}, \bibinfo {author} {\bibfnamefont {P.}~\bibnamefont {Bunyk}},
  \bibinfo {author} {\bibfnamefont {J.}~\bibnamefont {Raymond}},\ and\ \bibinfo
  {author} {\bibfnamefont {A.}~\bibnamefont {Roy}},\ }\bibfield  {title}
  {\bibinfo {title} {Next-{Generation} {Topology} of {D}-{Wave} {Quantum}
  {Processors}},\ }\href {https://doi.org/10.48550/arXiv.2003.00133} {\bibfield
   {journal} {\bibinfo  {journal} {arXiv:2003.00133}\ } (\bibinfo {year}
  {2020})}\BibitemShut {NoStop}%
\bibitem [{\citenamefont {Boothby}\ \emph {et~al.}(2021)\citenamefont
  {Boothby}, \citenamefont {King},\ and\ \citenamefont
  {Raymond}}]{boothby_zephyr_2021}%
  \BibitemOpen
  \bibfield  {author} {\bibinfo {author} {\bibfnamefont {K.}~\bibnamefont
  {Boothby}}, \bibinfo {author} {\bibfnamefont {A.~D.}\ \bibnamefont {King}},\
  and\ \bibinfo {author} {\bibfnamefont {J.}~\bibnamefont {Raymond}},\
  }\bibfield  {title} {\bibinfo {title} {Zephyr {Topology} of {D}-{Wave}
  {Quantum} {Processors}},\ }\href
  {https://www.dwavesys.com/media/2uznec4s/14-1056a-a_zephyr_topology_of_d-wave_quantum_processors.pdf}
  {\bibfield  {journal} {\bibinfo  {journal} {{Tech.} {Rep.} {D}-{Wave}}\ }
  (\bibinfo {year} {2021})}\BibitemShut {NoStop}%
\bibitem [{\citenamefont {Verstraete}\ \emph {et~al.}(2006)\citenamefont
  {Verstraete}, \citenamefont {Wolf}, \citenamefont {Perez-Garcia},\ and\
  \citenamefont {Cirac}}]{verstraete_criticality_2006}%
  \BibitemOpen
  \bibfield  {author} {\bibinfo {author} {\bibfnamefont {F.}~\bibnamefont
  {Verstraete}}, \bibinfo {author} {\bibfnamefont {M.~M.}\ \bibnamefont
  {Wolf}}, \bibinfo {author} {\bibfnamefont {D.}~\bibnamefont {Perez-Garcia}},\
  and\ \bibinfo {author} {\bibfnamefont {J.~I.}\ \bibnamefont {Cirac}},\
  }\bibfield  {title} {\bibinfo {title} {Criticality, the area law, and the
  computational power of projected entangled pair states},\ }\href
  {https://doi.org/10.1103/PhysRevLett.96.220601} {\bibfield  {journal}
  {\bibinfo  {journal} {Phys. Rev. Lett.}\ }\textbf {\bibinfo {volume} {96}},\
  \bibinfo {pages} {220601} (\bibinfo {year} {2006})}\BibitemShut {NoStop}%
\bibitem [{\citenamefont {Lubasch}\ \emph {et~al.}(2014)\citenamefont
  {Lubasch}, \citenamefont {Cirac},\ and\ \citenamefont
  {{Ba\~{n}uls}}}]{lubasch_unifying_2014}%
  \BibitemOpen
  \bibfield  {author} {\bibinfo {author} {\bibfnamefont {M.}~\bibnamefont
  {Lubasch}}, \bibinfo {author} {\bibfnamefont {J.~I.}\ \bibnamefont {Cirac}},\
  and\ \bibinfo {author} {\bibfnamefont {M.-C.}\ \bibnamefont {{Ba\~{n}uls}}},\
  }\bibfield  {title} {\bibinfo {title} {Unifying projected entangled pair
  state contractions},\ }\href
  {http://stacks.iop.org/1367-2630/16/i=3/a=033014} {\bibfield  {journal}
  {\bibinfo  {journal} {New J. Phys.}\ }\textbf {\bibinfo {volume} {16}},\
  \bibinfo {pages} {033014} (\bibinfo {year} {2014})}\BibitemShut {NoStop}%
\bibitem [{\citenamefont {Halko}\ \emph {et~al.}(2009)\citenamefont {Halko},
  \citenamefont {Martinsson},\ and\ \citenamefont
  {Tropp}}]{halko_finding_2009}%
  \BibitemOpen
  \bibfield  {author} {\bibinfo {author} {\bibfnamefont {N.}~\bibnamefont
  {Halko}}, \bibinfo {author} {\bibfnamefont {P.-G.}\ \bibnamefont
  {Martinsson}},\ and\ \bibinfo {author} {\bibfnamefont {J.~A.}\ \bibnamefont
  {Tropp}},\ }\bibfield  {title} {\bibinfo {title} {Finding structure with
  randomness: Probabilistic algorithms for constructing approximate matrix
  decompositions},\ }\href {https://api.semanticscholar.org/CorpusID:262495129}
  {\bibfield  {journal} {\bibinfo  {journal} {SIAM Rev.}\ }\textbf {\bibinfo
  {volume} {53}},\ \bibinfo {pages} {217} (\bibinfo {year} {2009})}\BibitemShut
  {NoStop}%
\bibitem [{\citenamefont {Martinsson}\ and\ \citenamefont
  {Tropp}(2020)}]{martinsson_randomized_2020}%
  \BibitemOpen
  \bibfield  {author} {\bibinfo {author} {\bibfnamefont {P.-G.}\ \bibnamefont
  {Martinsson}}\ and\ \bibinfo {author} {\bibfnamefont {J.~A.}\ \bibnamefont
  {Tropp}},\ }\bibfield  {title} {\bibinfo {title} {Randomized numerical linear
  algebra: Foundations and algorithms},\ }\href
  {https://api.semanticscholar.org/CorpusID:229167320} {\bibfield  {journal}
  {\bibinfo  {journal} {Acta Numer.}\ }\textbf {\bibinfo {volume} {29}},\
  \bibinfo {pages} {403 } (\bibinfo {year} {2020})}\BibitemShut {NoStop}%
\bibitem [{GPU()}]{GPU}%
  \BibitemOpen
  \href@noop {} {}\bibinfo {note} {Our numerical tests were performed on an
  Nvidia RTX3090/4090 GPUs with 24GB of memory.}\BibitemShut {Stop}%
\bibitem [{\citenamefont {Sinha}\ \emph {et~al.}(2024)\citenamefont {Sinha},
  \citenamefont {Rams},\ and\ \citenamefont
  {Dziarmaga}}]{sinha_efficient_2024}%
  \BibitemOpen
  \bibfield  {author} {\bibinfo {author} {\bibfnamefont {A.}~\bibnamefont
  {Sinha}}, \bibinfo {author} {\bibfnamefont {M.~M.}\ \bibnamefont {Rams}},\
  and\ \bibinfo {author} {\bibfnamefont {J.}~\bibnamefont {Dziarmaga}},\
  }\bibfield  {title} {\bibinfo {title} {Efficient representation of minimally
  entangled typical thermal states in two dimensions via projected entangled
  pair states},\ }\href {https://doi.org/10.1103/PhysRevB.109.045136}
  {\bibfield  {journal} {\bibinfo  {journal} {Phys. Rev. B}\ }\textbf {\bibinfo
  {volume} {109}},\ \bibinfo {pages} {045136} (\bibinfo {year}
  {2024})}\BibitemShut {NoStop}%
\bibitem [{\citenamefont {Golub}\ and\ \citenamefont
  {Van~Loan}(1996)}]{golub_matrix_2013}%
  \BibitemOpen
  \bibfield  {author} {\bibinfo {author} {\bibfnamefont {G.~H.}\ \bibnamefont
  {Golub}}\ and\ \bibinfo {author} {\bibfnamefont {C.~F.}\ \bibnamefont
  {Van~Loan}},\ }\href@noop {} {\emph {\bibinfo {title} {Matrix
  computations}}}\ (\bibinfo  {publisher} {The Johns Hopkins Press Ltd.,
  London},\ \bibinfo {year} {1996})\BibitemShut {NoStop}%
\bibitem [{\citenamefont {Pearl}(1982)}]{pearl_reverend_1982}%
  \BibitemOpen
  \bibfield  {author} {\bibinfo {author} {\bibfnamefont {J.}~\bibnamefont
  {Pearl}},\ }\bibfield  {title} {\bibinfo {title} {Reverend {Bayes} on
  inference engines: A distributed hierarchical approach},\ }in\ \href
  {https://api.semanticscholar.org/CorpusID:14936636} {\emph {\bibinfo
  {booktitle} {Probabilistic and Causal Inference}}}\ (\bibinfo  {publisher}
  {Association for Computing Machinery},\ \bibinfo {year} {1982})\BibitemShut
  {NoStop}%
\bibitem [{\citenamefont {Yedidia}\ \emph {et~al.}(2003)\citenamefont
  {Yedidia}, \citenamefont {Freeman},\ and\ \citenamefont
  {Weiss}}]{yedidia_understanding_2003}%
  \BibitemOpen
  \bibfield  {author} {\bibinfo {author} {\bibfnamefont {J.}~\bibnamefont
  {Yedidia}}, \bibinfo {author} {\bibfnamefont {W.}~\bibnamefont {Freeman}},\
  and\ \bibinfo {author} {\bibfnamefont {Y.}~\bibnamefont {Weiss}},\ }\bibfield
   {title} {\bibinfo {title} {Understanding belief propagation and its
  generalizations},\ }in\ \href {https://www.merl.com/publications/TR2001-22}
  {\emph {\bibinfo {booktitle} {Exploring Artificial Intelligence in the New
  Millennium}}},\ \bibinfo {editor} {edited by\ \bibinfo {editor}
  {\bibfnamefont {G.}~\bibnamefont {Lakemeyer}}\ and\ \bibinfo {editor}
  {\bibfnamefont {B.}~\bibnamefont {Nebel}}}\ (\bibinfo  {publisher} {Morgan
  Kaufmann Publishers},\ \bibinfo {year} {2003})\ Chap.~\bibinfo {chapter} {8},
  pp.\ \bibinfo {pages} {239--236}\BibitemShut {NoStop}%
\bibitem [{Note1()}]{Note1}%
  \BibitemOpen
  \bibinfo {note} {As some of the qubits and couplers that would form an ideal
  Pegasus/Zephyr lattice are not available in the quantum annealing hardware,
  we set the corresponding local fields and couplings to zero so that all
  solvers use the same instances}\BibitemShut {NoStop}%
\bibitem [{\citenamefont {Tasseff}\ \emph {et~al.}(2024)\citenamefont
  {Tasseff}, \citenamefont {Albash}, \citenamefont {Morrell}, \citenamefont
  {Vuffray}, \citenamefont {Lokhov}, \citenamefont {Misra},\ and\ \citenamefont
  {Coffrin}}]{tasseff_emerging_2022}%
  \BibitemOpen
  \bibfield  {author} {\bibinfo {author} {\bibfnamefont {B.}~\bibnamefont
  {Tasseff}}, \bibinfo {author} {\bibfnamefont {T.}~\bibnamefont {Albash}},
  \bibinfo {author} {\bibfnamefont {Z.}~\bibnamefont {Morrell}}, \bibinfo
  {author} {\bibfnamefont {M.}~\bibnamefont {Vuffray}}, \bibinfo {author}
  {\bibfnamefont {A.~Y.}\ \bibnamefont {Lokhov}}, \bibinfo {author}
  {\bibfnamefont {S.}~\bibnamefont {Misra}},\ and\ \bibinfo {author}
  {\bibfnamefont {C.}~\bibnamefont {Coffrin}},\ }\bibfield  {title} {\bibinfo
  {title} {On the emerging potential of quantum annealing hardware for
  combinatorial optimization},\ }\href
  {https://doi.org/10.1007/s10732-024-09530-5} {\bibfield  {journal} {\bibinfo
  {journal} {Journal of Heuristics}\ }\textbf {\bibinfo {volume} {30}},\
  \bibinfo {pages} {325–358} (\bibinfo {year} {2024})}\BibitemShut {NoStop}%
\bibitem [{\citenamefont {Perera}\ \emph {et~al.}(2020)\citenamefont {Perera},
  \citenamefont {Hamze}, \citenamefont {Raymond}, \citenamefont {Weigel},\ and\
  \citenamefont {Katzgraber}}]{perera_computational_2020}%
  \BibitemOpen
  \bibfield  {author} {\bibinfo {author} {\bibfnamefont {D.}~\bibnamefont
  {Perera}}, \bibinfo {author} {\bibfnamefont {F.}~\bibnamefont {Hamze}},
  \bibinfo {author} {\bibfnamefont {J.}~\bibnamefont {Raymond}}, \bibinfo
  {author} {\bibfnamefont {M.}~\bibnamefont {Weigel}},\ and\ \bibinfo {author}
  {\bibfnamefont {H.~G.}\ \bibnamefont {Katzgraber}},\ }\bibfield  {title}
  {\bibinfo {title} {Computational hardness of spin-glass problems with
  tile-planted solutions},\ }\href
  {https://doi.org/10.1103/PhysRevE.101.023316} {\bibfield  {journal} {\bibinfo
   {journal} {Phys. Rev. E}\ }\textbf {\bibinfo {volume} {101}},\ \bibinfo
  {pages} {023316} (\bibinfo {year} {2020})}\BibitemShut {NoStop}%
\bibitem [{\citenamefont {Perera}\ \emph {et~al.}(2021)\citenamefont {Perera},
  \citenamefont {Akpabio}, \citenamefont {Hamze}, \citenamefont {Mandra},
  \citenamefont {Rose}, \citenamefont {Aramon},\ and\ \citenamefont
  {Katzgraber}}]{perera_chook_2020}%
  \BibitemOpen
  \bibfield  {author} {\bibinfo {author} {\bibfnamefont {D.}~\bibnamefont
  {Perera}}, \bibinfo {author} {\bibfnamefont {I.}~\bibnamefont {Akpabio}},
  \bibinfo {author} {\bibfnamefont {F.}~\bibnamefont {Hamze}}, \bibinfo
  {author} {\bibfnamefont {S.}~\bibnamefont {Mandra}}, \bibinfo {author}
  {\bibfnamefont {N.}~\bibnamefont {Rose}}, \bibinfo {author} {\bibfnamefont
  {M.}~\bibnamefont {Aramon}},\ and\ \bibinfo {author} {\bibfnamefont {H.~G.}\
  \bibnamefont {Katzgraber}},\ }\bibfield  {title} {\bibinfo {title} {Chook --
  a comprehensive suite for generating binary optimization problems with
  planted solutions},\ }\href {https://doi.org/10.48550/arXiv.2005.14344}
  {\bibfield  {journal} {\bibinfo  {journal} {arXiv:2005.14344}\ } (\bibinfo
  {year} {2021})}\BibitemShut {NoStop}%
\bibitem [{\citenamefont {Vodeb}\ \emph {et~al.}(2024)\citenamefont {Vodeb},
  \citenamefont {Eržen}, \citenamefont {Hrga},\ and\ \citenamefont
  {Povh}}]{vodeb_accuracy_2024}%
  \BibitemOpen
  \bibfield  {author} {\bibinfo {author} {\bibfnamefont {J.}~\bibnamefont
  {Vodeb}}, \bibinfo {author} {\bibfnamefont {V.}~\bibnamefont {Eržen}},
  \bibinfo {author} {\bibfnamefont {T.}~\bibnamefont {Hrga}},\ and\ \bibinfo
  {author} {\bibfnamefont {J.}~\bibnamefont {Povh}},\ }\bibfield  {title}
  {\bibinfo {title} {Accuracy and performance evaluation of quantum, classical
  and hybrid solvers for the max-cut problem},\ }\href
  {https://doi.org/10.48550/arXiv.2412.07460} {\bibfield  {journal} {\bibinfo
  {journal} {arXiv:2412.07460}\ } (\bibinfo {year} {2024})}\BibitemShut
  {NoStop}%
\bibitem [{\citenamefont {Mohseni}\ \emph {et~al.}(2023)\citenamefont
  {Mohseni}, \citenamefont {Rams}, \citenamefont {Isakov}, \citenamefont
  {Eppens}, \citenamefont {Pielawa}, \citenamefont {Strumpfer}, \citenamefont
  {Boixo},\ and\ \citenamefont {Neven}}]{mohseni_sampling_2023}%
  \BibitemOpen
  \bibfield  {author} {\bibinfo {author} {\bibfnamefont {M.}~\bibnamefont
  {Mohseni}}, \bibinfo {author} {\bibfnamefont {M.~M.}\ \bibnamefont {Rams}},
  \bibinfo {author} {\bibfnamefont {S.~V.}\ \bibnamefont {Isakov}}, \bibinfo
  {author} {\bibfnamefont {D.}~\bibnamefont {Eppens}}, \bibinfo {author}
  {\bibfnamefont {S.}~\bibnamefont {Pielawa}}, \bibinfo {author} {\bibfnamefont
  {J.}~\bibnamefont {Strumpfer}}, \bibinfo {author} {\bibfnamefont
  {S.}~\bibnamefont {Boixo}},\ and\ \bibinfo {author} {\bibfnamefont
  {H.}~\bibnamefont {Neven}},\ }\bibfield  {title} {\bibinfo {title} {Sampling
  diverse near-optimal solutions via algorithmic quantum annealing},\ }\href
  {http://dx.doi.org/10.1103/PhysRevE.108.065303} {\bibfield  {journal}
  {\bibinfo  {journal} {Phys. Rev. E}\ }\textbf {\bibinfo {volume} {108}},\
  \bibinfo {pages} {065303} (\bibinfo {year} {2023})}\BibitemShut {NoStop}%
\bibitem [{\citenamefont {Zucca}\ \emph {et~al.}(2021)\citenamefont {Zucca},
  \citenamefont {Sadeghi}, \citenamefont {Mohseni},\ and\ \citenamefont
  {Amin}}]{zucca_diversity_2021}%
  \BibitemOpen
  \bibfield  {author} {\bibinfo {author} {\bibfnamefont {A.}~\bibnamefont
  {Zucca}}, \bibinfo {author} {\bibfnamefont {H.}~\bibnamefont {Sadeghi}},
  \bibinfo {author} {\bibfnamefont {M.}~\bibnamefont {Mohseni}},\ and\ \bibinfo
  {author} {\bibfnamefont {M.~H.}\ \bibnamefont {Amin}},\ }\bibfield  {title}
  {\bibinfo {title} {Diversity metric for evaluation of quantum annealing},\
  }\href {http://arxiv.org/abs/2110.10196} {\bibfield  {journal} {\bibinfo
  {journal} {arXiv:2110.10196}\ } (\bibinfo {year} {2021})}\BibitemShut
  {NoStop}%
\bibitem [{\citenamefont {Bomze}\ \emph {et~al.}(1999)\citenamefont {Bomze},
  \citenamefont {Budinich}, \citenamefont {Pardalos},\ and\ \citenamefont
  {Pelillo}}]{bomze_handbook_1999}%
  \BibitemOpen
  \bibfield  {author} {\bibinfo {author} {\bibfnamefont {I.~M.}\ \bibnamefont
  {Bomze}}, \bibinfo {author} {\bibfnamefont {M.}~\bibnamefont {Budinich}},
  \bibinfo {author} {\bibfnamefont {P.~M.}\ \bibnamefont {Pardalos}},\ and\
  \bibinfo {author} {\bibfnamefont {M.}~\bibnamefont {Pelillo}},\ }\bibinfo
  {title} {The maximum clique problem},\ in\ \href
  {https://doi.org/10.1007/978-1-4757-3023-4_1} {\emph {\bibinfo {booktitle}
  {Handbook of Combinatorial Optimization: Supplement Volume A}}},\ \bibinfo
  {editor} {edited by\ \bibinfo {editor} {\bibfnamefont {D.-Z.}\ \bibnamefont
  {Du}}\ and\ \bibinfo {editor} {\bibfnamefont {P.~M.}\ \bibnamefont
  {Pardalos}}}\ (\bibinfo  {publisher} {Springer US},\ \bibinfo {address}
  {Boston, MA},\ \bibinfo {year} {1999})\ pp.\ \bibinfo {pages}
  {1--74}\BibitemShut {NoStop}%
\bibitem [{Note2()}]{Note2}%
  \BibitemOpen
  \bibinfo {note} {We iterate over the list of states in a random order,
  building in the process a set of independent states. At each step of the
  iteration, a configuration is added to the list of independent states if it
  is independent of all configurations already in the list. We select the
  largest set obtained in all the restarts of that procedure. Typically, we
  observe saturation in obtained clique sizes after performing on the order of
  $100$ restarts.}\BibitemShut {Stop}%
\bibitem [{\citenamefont {Chen}\ \emph {et~al.}(2025)\citenamefont {Chen},
  \citenamefont {Jiang}, \citenamefont {Hangleiter},\ and\ \citenamefont
  {Schuch}}]{chen_sign_2024}%
  \BibitemOpen
  \bibfield  {author} {\bibinfo {author} {\bibfnamefont {J.}~\bibnamefont
  {Chen}}, \bibinfo {author} {\bibfnamefont {J.}~\bibnamefont {Jiang}},
  \bibinfo {author} {\bibfnamefont {D.}~\bibnamefont {Hangleiter}},\ and\
  \bibinfo {author} {\bibfnamefont {N.}~\bibnamefont {Schuch}},\ }\bibfield
  {title} {\bibinfo {title} {Sign problem in tensor-network contraction},\
  }\href {http://dx.doi.org/10.1103/PRXQuantum.6.010312} {\bibfield  {journal}
  {\bibinfo  {journal} {PRX Quantum}\ }\textbf {\bibinfo {volume} {6}},\
  \bibinfo {pages} {010312} (\bibinfo {year} {2025})}\BibitemShut {NoStop}%
\bibitem [{\citenamefont {Mohseni}\ \emph {et~al.}(2024)\citenamefont
  {Mohseni}, \citenamefont {Scherer}, \citenamefont {Johnson}, \citenamefont
  {Wertheim}, \citenamefont {Otten}, \citenamefont {Aadit}, \citenamefont
  {Alexeev}, \citenamefont {Bresniker}, \citenamefont {Camsari}, \citenamefont
  {Chapman}, \citenamefont {Chatterjee}, \citenamefont {Dagnew}, \citenamefont
  {Esposito}, \citenamefont {Fahim}, \citenamefont {Fiorentino}, \citenamefont
  {Gajjar}, \citenamefont {Khalid}, \citenamefont {Kong}, \citenamefont
  {Kulchytskyy}, \citenamefont {Kyoseva}, \citenamefont {Li}, \citenamefont
  {Lott}, \citenamefont {Markov}, \citenamefont {McDermott}, \citenamefont
  {Pedretti}, \citenamefont {Rao}, \citenamefont {Rieffel}, \citenamefont
  {Silva}, \citenamefont {Sorebo}, \citenamefont {Spentzouris}, \citenamefont
  {Steiner}, \citenamefont {Torosov}, \citenamefont {Venturelli}, \citenamefont
  {Visser}, \citenamefont {Webb}, \citenamefont {Zhan}, \citenamefont {Cohen},
  \citenamefont {Ronagh}, \citenamefont {Ho}, \citenamefont {Beausoleil},\ and\
  \citenamefont {Martinis}}]{Mohseni2024}%
  \BibitemOpen
  \bibfield  {author} {\bibinfo {author} {\bibfnamefont {M.}~\bibnamefont
  {Mohseni}}, \bibinfo {author} {\bibfnamefont {A.}~\bibnamefont {Scherer}},
  \bibinfo {author} {\bibfnamefont {K.~G.}\ \bibnamefont {Johnson}}, \bibinfo
  {author} {\bibfnamefont {O.}~\bibnamefont {Wertheim}}, \bibinfo {author}
  {\bibfnamefont {M.}~\bibnamefont {Otten}}, \bibinfo {author} {\bibfnamefont
  {N.~A.}\ \bibnamefont {Aadit}}, \bibinfo {author} {\bibfnamefont
  {Y.}~\bibnamefont {Alexeev}}, \bibinfo {author} {\bibfnamefont {K.~M.}\
  \bibnamefont {Bresniker}}, \bibinfo {author} {\bibfnamefont {K.~Y.}\
  \bibnamefont {Camsari}}, \bibinfo {author} {\bibfnamefont {B.}~\bibnamefont
  {Chapman}}, \bibinfo {author} {\bibfnamefont {S.}~\bibnamefont {Chatterjee}},
  \bibinfo {author} {\bibfnamefont {G.~A.}\ \bibnamefont {Dagnew}}, \bibinfo
  {author} {\bibfnamefont {A.}~\bibnamefont {Esposito}}, \bibinfo {author}
  {\bibfnamefont {F.}~\bibnamefont {Fahim}}, \bibinfo {author} {\bibfnamefont
  {M.}~\bibnamefont {Fiorentino}}, \bibinfo {author} {\bibfnamefont
  {A.}~\bibnamefont {Gajjar}}, \bibinfo {author} {\bibfnamefont
  {A.}~\bibnamefont {Khalid}}, \bibinfo {author} {\bibfnamefont
  {X.}~\bibnamefont {Kong}}, \bibinfo {author} {\bibfnamefont {B.}~\bibnamefont
  {Kulchytskyy}}, \bibinfo {author} {\bibfnamefont {E.}~\bibnamefont
  {Kyoseva}}, \bibinfo {author} {\bibfnamefont {R.}~\bibnamefont {Li}},
  \bibinfo {author} {\bibfnamefont {P.~A.}\ \bibnamefont {Lott}}, \bibinfo
  {author} {\bibfnamefont {I.~L.}\ \bibnamefont {Markov}}, \bibinfo {author}
  {\bibfnamefont {R.~F.}\ \bibnamefont {McDermott}}, \bibinfo {author}
  {\bibfnamefont {G.}~\bibnamefont {Pedretti}}, \bibinfo {author}
  {\bibfnamefont {P.}~\bibnamefont {Rao}}, \bibinfo {author} {\bibfnamefont
  {E.}~\bibnamefont {Rieffel}}, \bibinfo {author} {\bibfnamefont
  {A.}~\bibnamefont {Silva}}, \bibinfo {author} {\bibfnamefont
  {J.}~\bibnamefont {Sorebo}}, \bibinfo {author} {\bibfnamefont
  {P.}~\bibnamefont {Spentzouris}}, \bibinfo {author} {\bibfnamefont
  {Z.}~\bibnamefont {Steiner}}, \bibinfo {author} {\bibfnamefont
  {B.}~\bibnamefont {Torosov}}, \bibinfo {author} {\bibfnamefont
  {D.}~\bibnamefont {Venturelli}}, \bibinfo {author} {\bibfnamefont {R.~J.}\
  \bibnamefont {Visser}}, \bibinfo {author} {\bibfnamefont {Z.}~\bibnamefont
  {Webb}}, \bibinfo {author} {\bibfnamefont {X.}~\bibnamefont {Zhan}}, \bibinfo
  {author} {\bibfnamefont {Y.}~\bibnamefont {Cohen}}, \bibinfo {author}
  {\bibfnamefont {P.}~\bibnamefont {Ronagh}}, \bibinfo {author} {\bibfnamefont
  {A.}~\bibnamefont {Ho}}, \bibinfo {author} {\bibfnamefont {R.~G.}\
  \bibnamefont {Beausoleil}},\ and\ \bibinfo {author} {\bibfnamefont {J.~M.}\
  \bibnamefont {Martinis}},\ }\bibfield  {title} {\bibinfo {title} {How to
  build a quantum supercomputer: Scaling from hundreds to millions of qubits},\
  }\href {https://doi.org/10.48550/arXiv.2411.10406} {\bibfield  {journal}
  {\bibinfo  {journal} {arXiv:2411.10406}\ } (\bibinfo {year}
  {2024})}\BibitemShut {NoStop}%
\bibitem [{\citenamefont {Pedretti}\ \emph {et~al.}(2023)\citenamefont
  {Pedretti}, \citenamefont {Böhm}, \citenamefont {Hizzani}, \citenamefont
  {Bhattacharya}, \citenamefont {Bruel}, \citenamefont {Moon}, \citenamefont
  {Serebryakov}, \citenamefont {Strukov}, \citenamefont {Strachan},
  \citenamefont {Ignowski} \emph {et~al.}}]{pedretti2023zeroth}%
  \BibitemOpen
  \bibfield  {author} {\bibinfo {author} {\bibfnamefont {G.}~\bibnamefont
  {Pedretti}}, \bibinfo {author} {\bibfnamefont {F.}~\bibnamefont {Böhm}},
  \bibinfo {author} {\bibfnamefont {M.}~\bibnamefont {Hizzani}}, \bibinfo
  {author} {\bibfnamefont {T.}~\bibnamefont {Bhattacharya}}, \bibinfo {author}
  {\bibfnamefont {P.}~\bibnamefont {Bruel}}, \bibinfo {author} {\bibfnamefont
  {J.}~\bibnamefont {Moon}}, \bibinfo {author} {\bibfnamefont {S.}~\bibnamefont
  {Serebryakov}}, \bibinfo {author} {\bibfnamefont {D.}~\bibnamefont
  {Strukov}}, \bibinfo {author} {\bibfnamefont {J.}~\bibnamefont {Strachan}},
  \bibinfo {author} {\bibfnamefont {J.}~\bibnamefont {Ignowski}}, \emph
  {et~al.},\ }\bibfield  {title} {\bibinfo {title} {Zeroth and higher-order
  logic with content addressable memories},\ }in\ \href
  {https://doi.org/10.1109/IEDM45741.2023.10413853} {\emph {\bibinfo
  {booktitle} {2023 International Electron Devices Meeting (IEDM)}}}\ (\bibinfo
  {year} {2023})\ pp.\ \bibinfo {pages} {1--4}\BibitemShut {NoStop}%
\bibitem [{\citenamefont {Chowdhury}\ \emph {et~al.}(2023)\citenamefont
  {Chowdhury}, \citenamefont {Grimaldi}, \citenamefont {Aadit}, \citenamefont
  {Niazi}, \citenamefont {Mohseni}, \citenamefont {Kanai}, \citenamefont
  {Ohno}, \citenamefont {Fukami}, \citenamefont {Theogarajan}, \citenamefont
  {Finocchio}, \citenamefont {Datta},\ and\ \citenamefont
  {Camsari}}]{chowdhury2023fullstack}%
  \BibitemOpen
  \bibfield  {author} {\bibinfo {author} {\bibfnamefont {S.}~\bibnamefont
  {Chowdhury}}, \bibinfo {author} {\bibfnamefont {A.}~\bibnamefont {Grimaldi}},
  \bibinfo {author} {\bibfnamefont {N.~A.}\ \bibnamefont {Aadit}}, \bibinfo
  {author} {\bibfnamefont {S.}~\bibnamefont {Niazi}}, \bibinfo {author}
  {\bibfnamefont {M.}~\bibnamefont {Mohseni}}, \bibinfo {author} {\bibfnamefont
  {S.}~\bibnamefont {Kanai}}, \bibinfo {author} {\bibfnamefont
  {H.}~\bibnamefont {Ohno}}, \bibinfo {author} {\bibfnamefont {S.}~\bibnamefont
  {Fukami}}, \bibinfo {author} {\bibfnamefont {L.}~\bibnamefont {Theogarajan}},
  \bibinfo {author} {\bibfnamefont {G.}~\bibnamefont {Finocchio}}, \bibinfo
  {author} {\bibfnamefont {S.}~\bibnamefont {Datta}},\ and\ \bibinfo {author}
  {\bibfnamefont {K.~Y.}\ \bibnamefont {Camsari}},\ }\bibfield  {title}
  {\bibinfo {title} {A full-stack view of probabilistic computing with p-bits:
  Devices, architectures, and algorithms},\ }\href
  {https://doi.org/10.1109/JXCDC.2023.3256981} {\bibfield  {journal} {\bibinfo
  {journal} {IEEE Journal on Exploratory Solid-State Computational Devices and
  Circuits}\ }\textbf {\bibinfo {volume} {9}},\ \bibinfo {pages} {1} (\bibinfo
  {year} {2023})}\BibitemShut {NoStop}%
\bibitem [{\citenamefont {\'{S}miechrzalski}\ \emph {et~al.}(2025)\citenamefont
  {\'{S}miechrzalski}, \citenamefont {Dziubyna}, \citenamefont {Ja\l{}owiecki},
  \citenamefont {Mzaouali}, \citenamefont {Pawela}, \citenamefont {Gardas},\
  and\ \citenamefont {Rams}}]{SGPEPS}%
  \BibitemOpen
  \bibfield  {author} {\bibinfo {author} {\bibfnamefont {T.}~\bibnamefont
  {\'{S}miechrzalski}}, \bibinfo {author} {\bibfnamefont {A.~M.}\ \bibnamefont
  {Dziubyna}}, \bibinfo {author} {\bibfnamefont {K.}~\bibnamefont
  {Ja\l{}owiecki}}, \bibinfo {author} {\bibfnamefont {Z.}~\bibnamefont
  {Mzaouali}}, \bibinfo {author} {\bibfnamefont {{\L}.}~\bibnamefont {Pawela}},
  \bibinfo {author} {\bibfnamefont {B.}~\bibnamefont {Gardas}},\ and\ \bibinfo
  {author} {\bibfnamefont {M.~M.}\ \bibnamefont {Rams}},\ }\bibfield  {title}
  {\bibinfo {title} {Spinglasspeps.jl: Tensor-network package for ising-like
  optimization on quasi-two-dimensional graphs},\ }\href
  {https://doi.org/10.48550/arXiv.2411.10406} {\bibfield  {journal} {\bibinfo
  {journal} {arXiv:2502.02317}\ } (\bibinfo {year} {2025})},\ \bibinfo {note}
  {\url{https://github.com/euro-hpc-pl/SpinGlassPEPS.jl}}\BibitemShut {NoStop}%
\bibitem [{dat()}]{database}%
  \BibitemOpen
  \href@noop {} {}\bibinfo {note}
  {\url{https://doi.org/10.18150/OUGF1O}}\BibitemShut {NoStop}%
\end{thebibliography}%

\appendix
\onecolumngrid
\setcounter{equation}{0}
\setcounter{figure}{0}
\renewcommand{\theequation}{A\arabic{equation}}
\renewcommand{\thefigure}{A\arabic{figure}}

\newpage

\section{Auxiliary results}
\label{sec:aux_results}
\begin{figure*}[h!]
\begin{centering}
\includegraphics[width=0.99\textwidth]{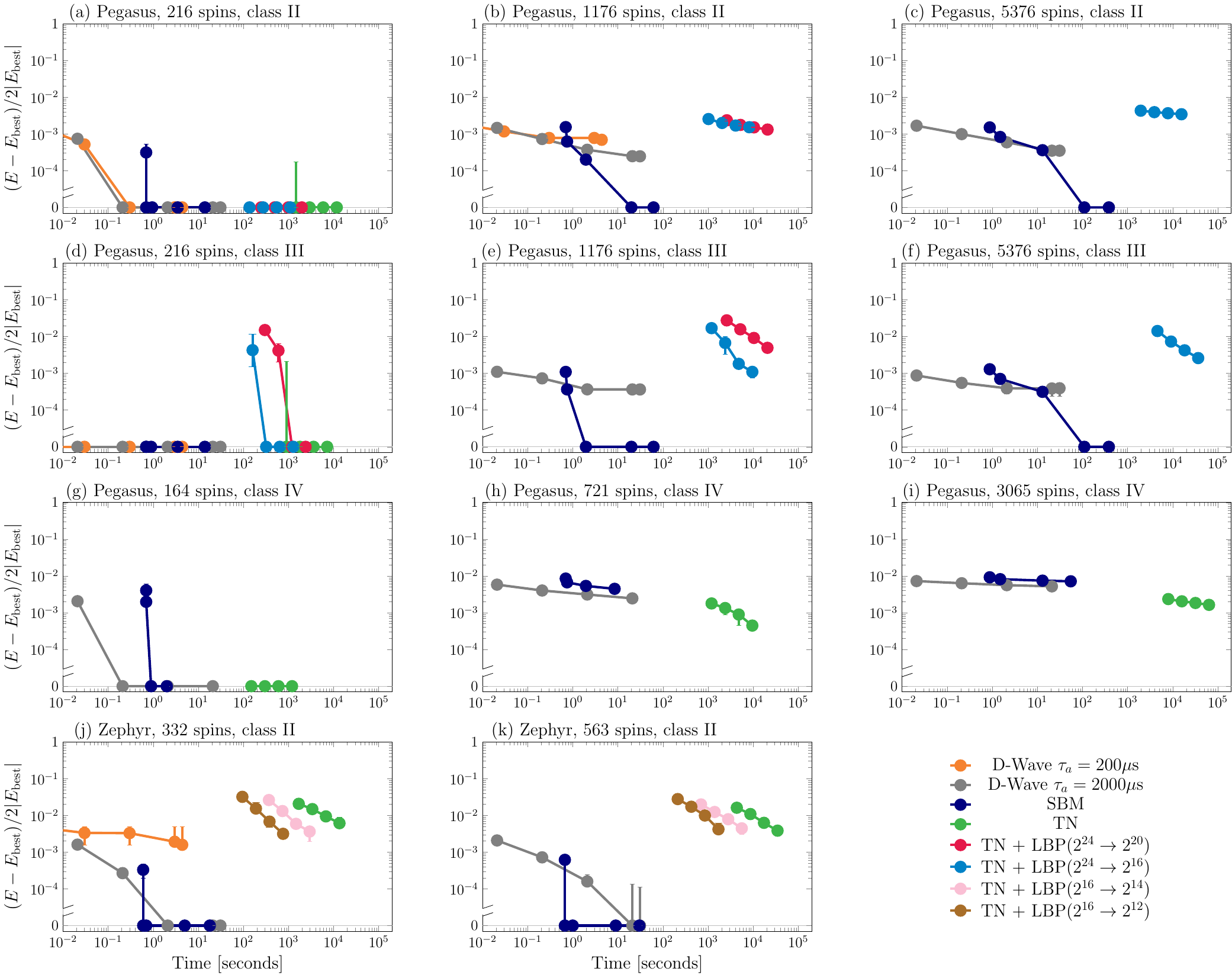}
\par\end{centering}
\caption{Time to approximation ratio. Here, we show the results for instances in class II, III and VI for Pegasus and Zephyr geometries for a selection of system sizes. The data complement the results in Fig.~\ref{fig:10} and were obtained with the same simulation parameters. We show the median based on 20 instances.}
\label{fig:A1}
\end{figure*}

In Fig.~\ref{fig:A1}, we show time to approximation ratio for class II, III and IV instances achieved by all examined solvers: D-Wave QA, SBM and TN approach. {In contrast to classes I, II, and III, which exhibit similar performance trends, class IV instances demonstrate an advantage in energy results achieved by the TN approach. Although TN outperforms both D-Wave QA and SBM in the terms of energy for instances with 721 and 3065 spins, it fails to identify their ground states.} 

\begin{figure*}[t]
\begin{centering}
\includegraphics[width=0.99\textwidth]{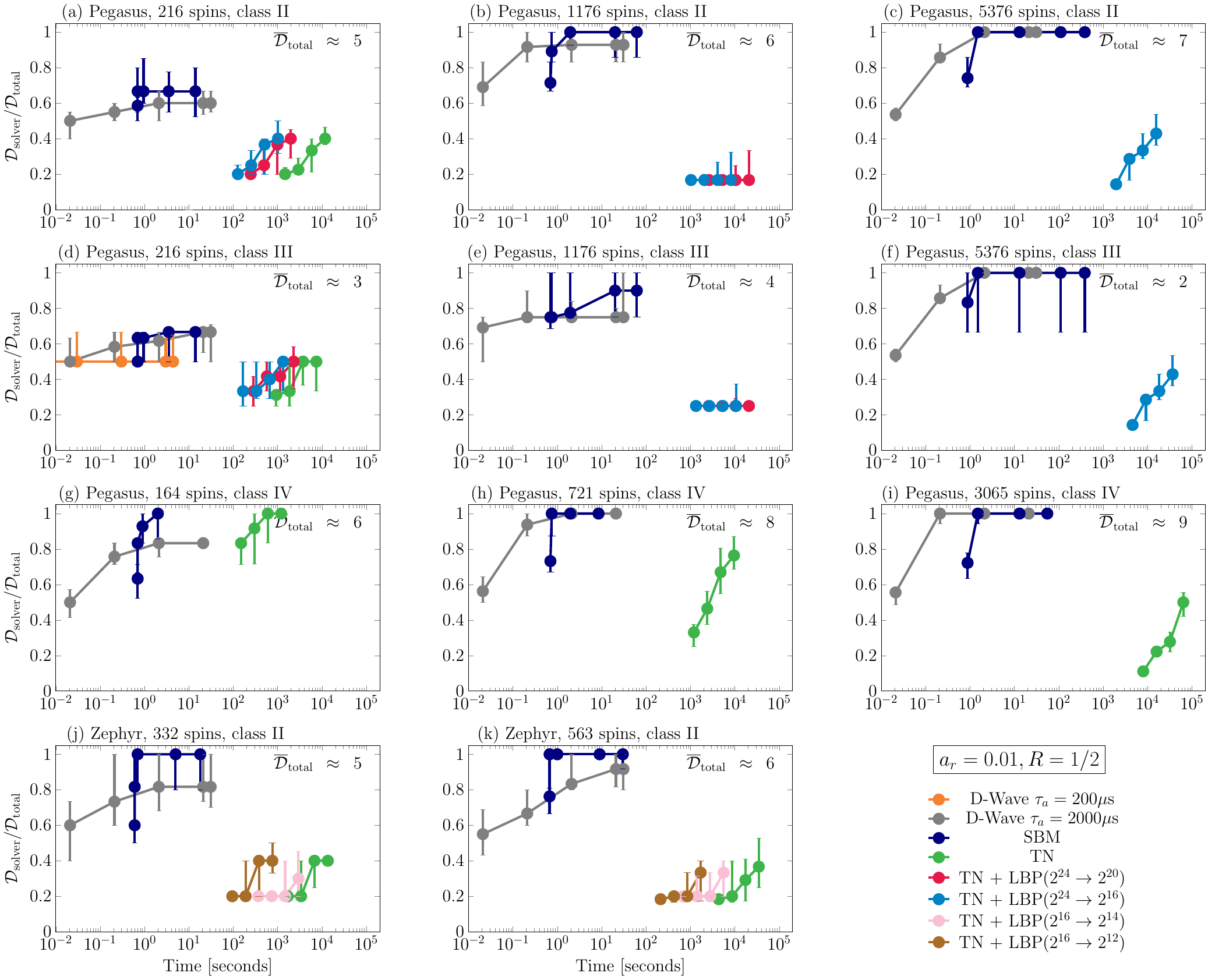}
\par\end{centering}
\caption{Time to diversity ratio for $R=1/2$.  We show the time needed to reach a given fraction of diverse solutions within $a_r=0.01$ and $R=1/2$ for instance class II, III and IV, complementing the data in Fig.~\ref{fig:11}.
}
\label{fig:A2}
\end{figure*}

\begin{figure*}[t]
\begin{centering}
\includegraphics[width=0.99\textwidth]{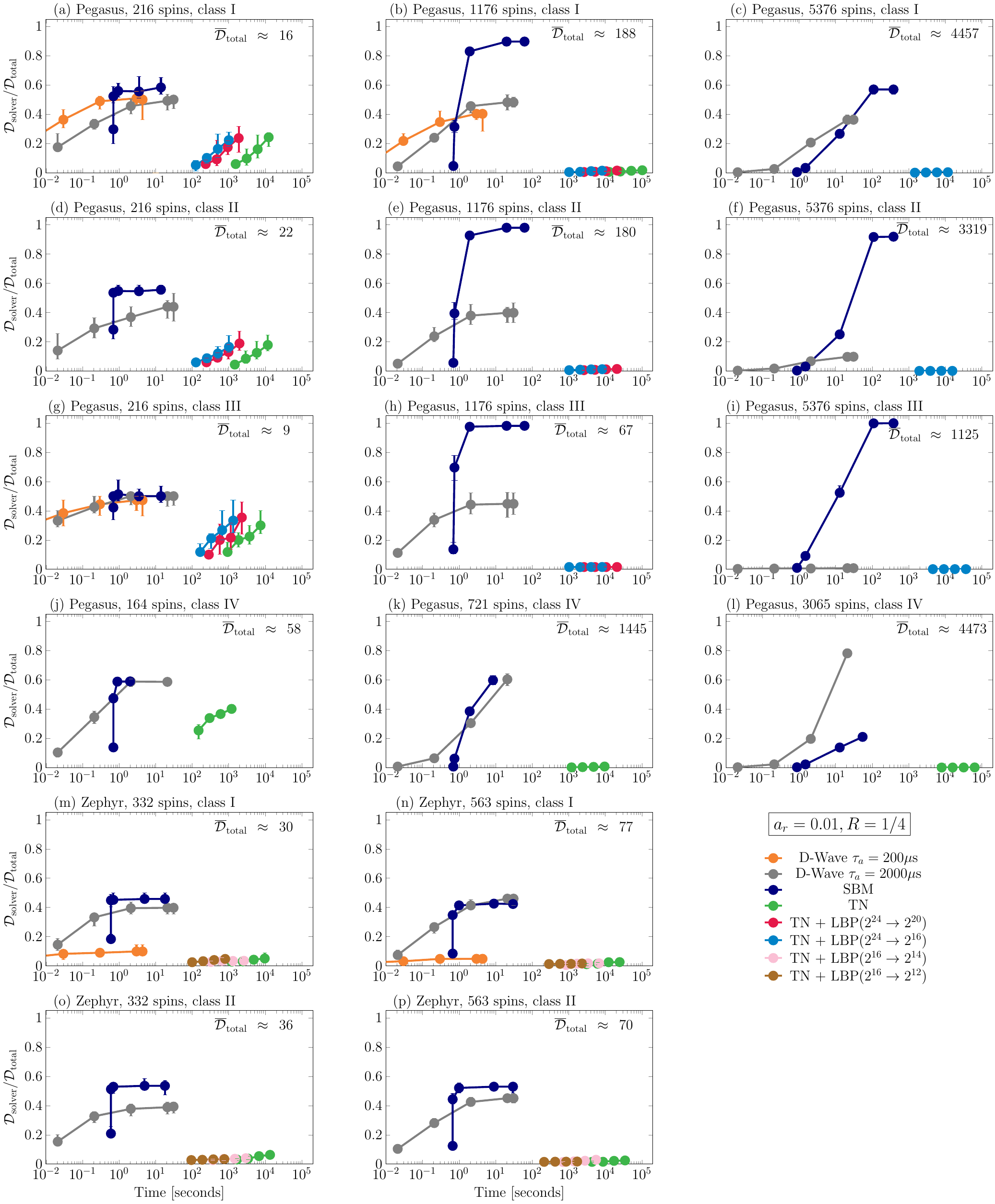}
\par\end{centering}
\caption{Time to diversity ratio for $R=1/4$. We show the data for all four instance classes and approximation ratio $a_r=0.01$, complementing similar results in Figs.~\ref{fig:11} and~\ref{fig:A2} for $R=1/2$. For large system sizes, the fraction of diverse solutions identified by TN becomes negligible as compared to two other solvers, where a limited available width of a branch and bound procedure $M=1024$, apparently, allows identifying only a small fraction of independent solutions for $\mathcal{D}_{\rm total}$ in hundreds. The simulation parameters are the same as in Fig.~\ref{fig:10}.
}
\label{fig:A3}
\end{figure*}

Time to diversity ratio for classes I, II, III and IV with the distance threshold $R=1/2$ and $R=1/4$ are shown in the Fig.~\ref{fig:A2} and~\ref{fig:A3}, respectively. 

\clearpage
\newpage

\clearpage
\newpage

\section{Square lattice}
\label{sec:square}
\begin{figure*}[h!]
\begin{centering}
\includegraphics[width=0.80\textwidth]{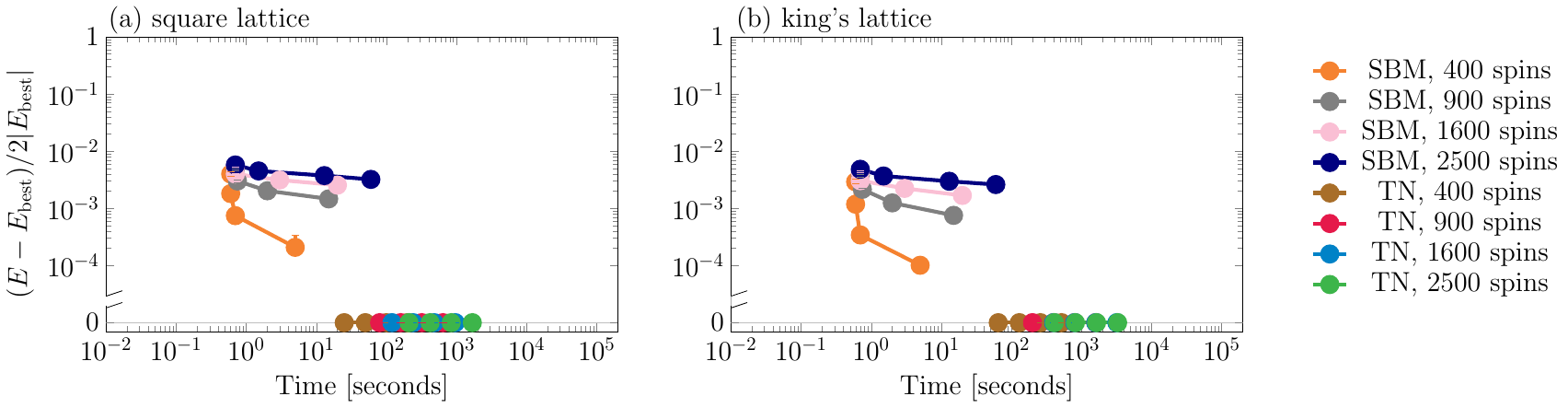}
\par\end{centering}
\caption{Time to approximation ratio for class I instances on a square lattice, panel (a), and a king's lattice (square lattice with diagonals) in (b) for a selection of system sizes. Like in Fig.~\ref{fig:10}, we show here the median based on 20 problem instances. The best energies $E_{\rm{best}}$ have been chosen from the results of SBM and TN algorithms. Here for TN, we show data for bond dimension $\chi = 16$, maximum number of states considered during the search set to $M=256$ and inverse temperature $\beta = 4.0$.}
\label{fig:A4}
\end{figure*}

To illustrate our motivations to utilize TN in optimization problems, we present results for class I instances on a square lattice with one spin per lattice site and king's lattice generated by adding diagonals to the square lattice. For reference results, we conducted simulations using SBM. The results focusing on best energies are presented in Fig.~\ref{fig:A4}. For those geometries, the TN-based technique consistently finds states with lower energies than SBM. King's lattice proves to be harder for TN than a square lattice. Indeed, for a square lattice, nearly all runs for all instances and lattice transformations for $\beta=4$ and $\beta=8$ result in the ground state (beyond the median for $\beta=4$ shown in Fig.~\ref{fig:A4}). For the king's lattice, the success rate in identifying the ground state falls in a  $50\times50$ lattice to around $90\%$ for $\beta=4$, and around $50\%$ for $\beta=8$. 

\begin{figure*}[b]
\begin{centering}
\includegraphics[width=0.70\textwidth]{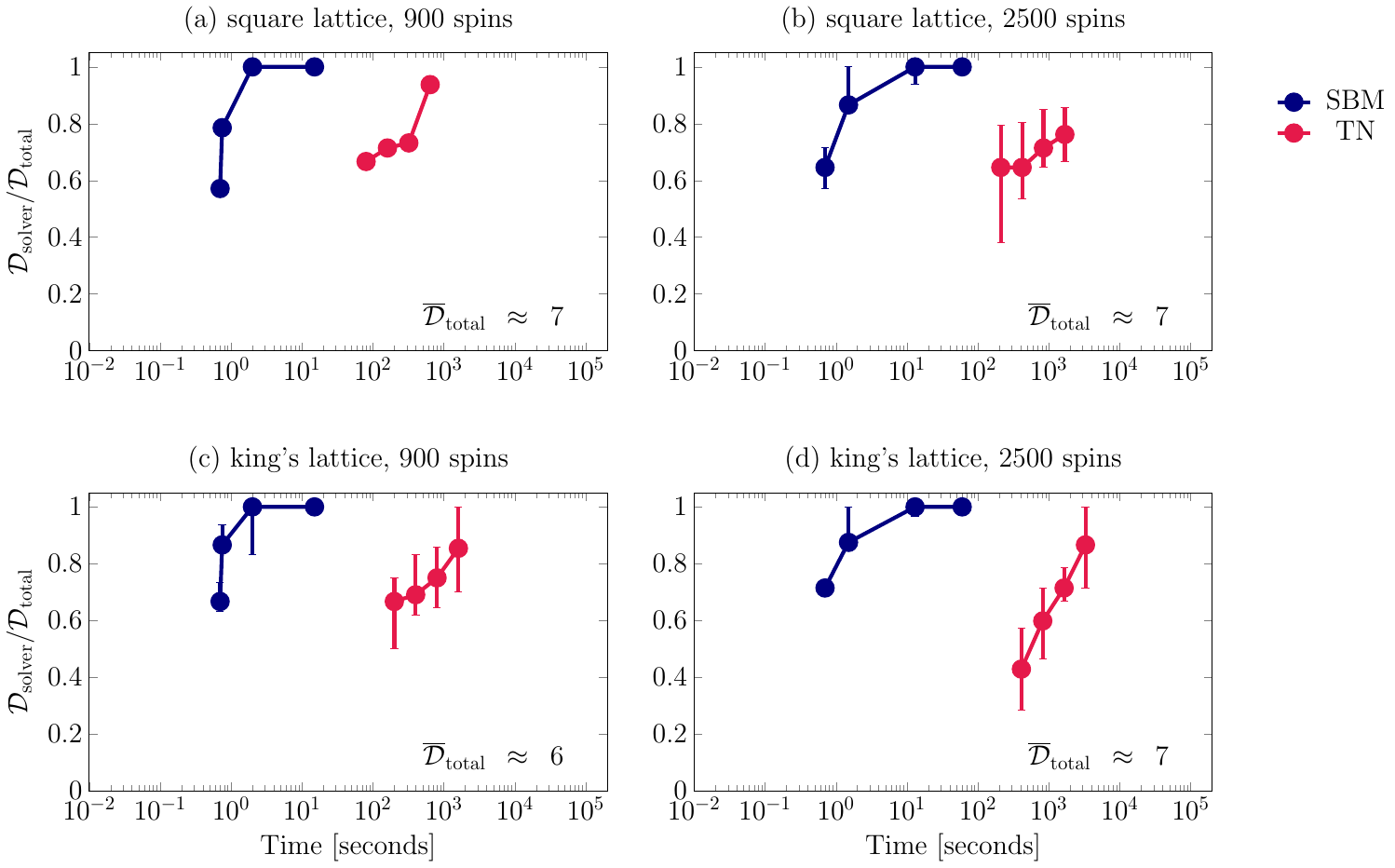}
\par\end{centering}
\caption{Time to diversity ratio for class I instances on a square lattice, panels (a) and (b), and king's lattice in panels (c) and (d) for the distance threshold $R=1/2$ and approximation ratio $a_r = 0.005$. The simulation parameters are the same as in Fig.~\ref{fig:A5}.}
\label{fig:A5}
\end{figure*}

The diversity of solutions (vastly different, with $R=1/2$) found by both solvers is comparable, see Fig.~\ref{fig:A5}, indicating that both solvers are capable of sampling from the same distant parts of the low-energy manifolds. However, TN does not have an advantage in that metric anymore.

Finally, we have also tested the zipper boundary MPS contraction procedure using randomized SVD, see Sec.~\ref{sec:zipper}, against a more standard procedure where the entire MPO (related to a PEPS row) is applied first to the previous boundary-MPS, the result is truncated using full-rank SVD, and fine-tuned variationally. The quality of the results obtained from both procedures turns out to be similar in those problems. 


\clearpage
\newpage

\section{Decomposition of PEPS into MPOs, optimal $\beta$ and stability test}
\label{stability}
\begin{figure*}[h]
\begin{centering}
\includegraphics[width=0.63\textwidth]{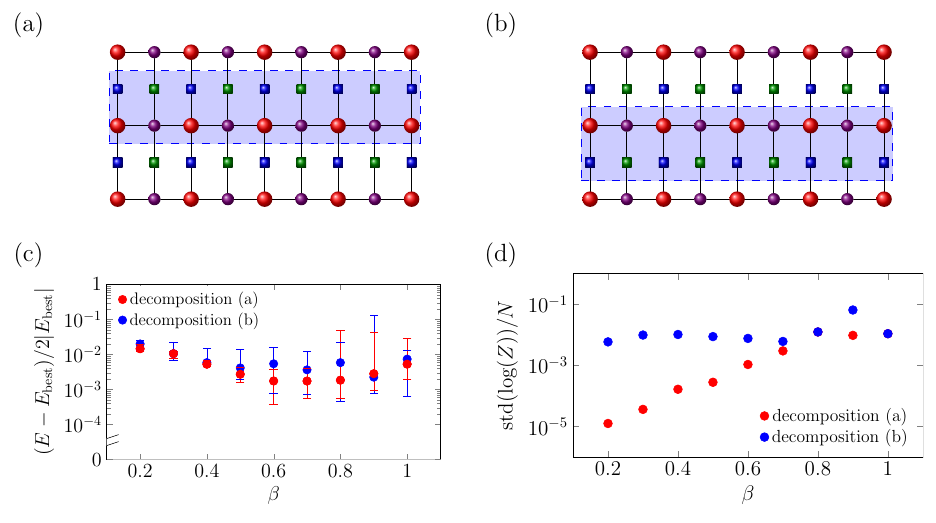}
\par\end{centering}
\caption{PEPS tensor network can be decomposed into MPOs in many ways, e.g., once shown in panels (a) and (b) with pale-blue boxes. In panels (c) and (d), we focus on a single instance in class I for the Pegasus graph with 1176 spins. Panel (c) displays the energy obtained by the TN solver above the best energy identified for this instance. For each $\beta$ and decomposition, we show the median from 8 runs of the algorithm using different rotations of the lattice (effectively, performing the search and the contraction from various corners of the lattice). The best and worst results are indicated by error bars. Panel (d) focuses on the stability of contraction, using the partition function as the metric, comparing the results obtained for 8 lattice rotations.}
\label{fig:A6}
\end{figure*}

We can consider various types of decomposition of the PEPS network into MPOs, as shown in Fig.~\ref{fig:A6}(a) and (b) with pale-blue boxes. All the results presented in this article employ the decomposition in (a).
Here, we compare the influence of such decomposition using a single instance in class I on the Pegasus graph with 1176 spins as an example. At the same time, we probe various inverse temperatures $\beta$. In Fig.~\ref{fig:A6}(c), we show the relative energy above the ground state for runs using all 8 lattice rotations. We observe a weak sensitivity of the results both on the decomposition and on $\beta$. All our results in the main text have been using $\beta=0.5$ for the Pegasus graph. Based on a similar analysis, we decided on $\beta=1.0$ for instances defined on the Zephyr graph.
In Fig.~\ref{fig:A6}(d), we show another metric indicating the contraction stability. Namely, we compare the value of the partition function $Z$ following 8 lattice rotations (hence, performing boundary-MPS contraction of the PEPS network from various edges of the lattice). Panel (d) shows the standard deviation among the 8 obtained values for each $\beta$ and decomposition. Decomposition (a) seems more stable within this metric for the lowest values of $\beta$, with the deviation growing systematically as $\beta$ increases.

\begin{figure*}[b]
\begin{centering}
\includegraphics[width=0.73\textwidth]{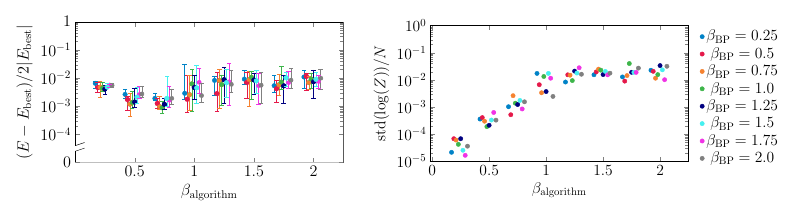}
\par\end{centering}
\caption{Influence of inverse temperature on the stability of the results utilizing LBP for dimensional reduction, where we independently change $\beta_{\rm{BP}}$ in LBP and $\beta_{\rm{algorithm}}$ in TN branch-and-bound algorithm. We consider the same single instance and the same performance metrics as in Fig.~\ref{fig:A6}(c,d). We utilize LBP to truncate the number of states in each node from $2^{24}$ to $2^{16}$. For better visibility of the data on the graph, we spaced out the data for different temperatures $\beta_{\rm{BP}}$.}
\label{fig:A7}
\end{figure*}

Finally, we focus on dimensional reduction based on LBP and how the results depend on $\beta$. The data are presented in Fig.~\ref{fig:A7}, where we perform simulations for the same single instance as in Fig.~\ref{fig:A6} above. In principle, the LBP procedure and TN branch-and-bound algorithm do not have to employ the same $\beta$, and here we probe various combinations of $\beta_{\rm{BP}}$ and $\beta_{\rm{algorithm}}$ (in the main article, they are set at the same value). The overall consistency of the results seems mostly influenced by $\beta_{\rm{algorithm}}$.

\newpage

\section{Simulated Bifurcation Machine}
\label{sec:sbm}

Building upon the concept of quantum adiabatic optimization using a network of nonlinear quantum oscillators, the Simulated Bifurcation Machine (SBM) leverages the phenomenon of bifurcation within a classical nonlinear system governed by Hamilton's equations of motion~\cite{goto_combinatorial_2019}:

\begin{equation}
\begin{split}
    \dot{q}_i &= \frac{\partial H_{\text{cl}}}{\partial p_i} = a_0 p_i \\
    \dot{p}_i &= -\frac{\partial H_{\text{cl}}}{\partial q_i} = -\left[q_i^2 + a_0 - a(t)\right] q_i + c_0 \Phi(q_i). 
\end{split}
\label{eq:sbm}
\end{equation}
where $\Phi(q_i) = \sum_{j=1}^{N} J_{ij} q_j + h_i$, and 
\begin{equation}
    H_{\text{cl}} = \frac{a_0}{2} \sum_{i=1}^{N} p_i^2 + \sum_{i=1}^{N} \left(\frac{q_i^4}{4} + \frac{a_0 - a(t)}{2} q_i^2 \right) 
      - c_0
    \sum_{i=1}^{N} \left(h_i q_i + \frac{1}{2}\sum_{j=1}^N J_{ij} q_i q_j\right), 
    \label{eq:clHam}
\end{equation}
where \(J\), \(h\) are defined in Eq.~\eqref{eq:Ising}.

This system can be interpreted as particles with mass \(1/a_0\), positions \(q_i\), and momenta \(p_i\), moving within a time-dependent potential and interacting through Ising-like couplings. The constants \(a_0\) and \(c_0\) are hyperparameters (in this work we set \(a_0 = 1\) and \(c_0 = 1/\lambda_{\text{max}}\), where \(\lambda_{\text{max}}\) denotes the largest eigenvalue of \(J\)), while \(a(t)\) is a continuous (here linear) increasing function of time that drives the system across a bifurcation point occurring approximately when \(a(t) = a_0\). Beyond that point, the energy landscape governing systems' evolution approximately encodes the \emph{local} minima of the Ising term, driving particles to move toward low-energy solutions. The final (discrete) solutions are then obtained as \(s_i = f(q_i^s)\), where $q_i^s$ is the steady state of~\eqref{eq:sbm}, and $f$ is taken as a sign function, for simplicity. 

To mitigate errors resulting from the relaxation of binary variables to continuous ones, a modified version of SBM (which was used in this work) was proposed in~\cite{goto_highperformance_2021}. Therein, the nonlinear term, \(q_i^4\), was replaced with perfectly inelastic walls at \(|q_i| = 1\), and the Ising contribution to the dynamics is discretized. Specifically, the term \(\sum_{j=1}^N J_{ij} q_j\) in Eq.~\eqref{eq:sbm} is substituted with \(\sum_{j=1}^N J_{ij} \mathrm{sign}(q_j)\). These adjustments ensure that the dynamics can be halted when \(a(t) = a_0\), with the system residing in a local minimum. In general, there is no guarantee that $f$ maps $q_i^s$ onto the ground state of~\eqref{eq:Ising}, $s_i^g$. Finding a function $f$ (if such exists) that ensures this mapping remains an open problem.

The differential equations~\eqref{eq:sbm} are separable with respect to positions, $q_i$ and momenta, $p_i$. This allows to integrate them numerically using simple yet stable symplectic Euler method, which can also be implemented on GPUs efficiently. Furthermore, the chaotic nature of the SBM equations implies sensitivity to initial conditions. Since each simulation is independent, multiple (random) starting points can be evolved simultaneously, enhancing efficiency and enabling extensive parallelization.

\newpage

\section{Quantum annealing devices}
D-Wave reference results for the Pegasus instances were obtained using the Advantage\_system6.1, while Zephyr instances were solved on the Advantage2\_prototype1.1. The properties of both machines are detailed in Table~\ref{machine_prop}.
\begin{table}[h!]
\centering
\caption{Physical properties of the D-Wave Advantage\_system6.1 machine and Advantage2\_prototype1.1}
\begin{tabular}{ | l|l|l|}
 \hline
\textbf{Parameter} & \textbf{Advantage\_system6.1} & \textbf{Advantage2\_prototype1.1} \\ 
 \hline
 \hline
Qubits & $5616$ & $563$   \\  
\hline
 Couplers & $40135$  & $4790$   \\
 \hline
 Qubit temperature (mK) & $16.0 \pm 0.1$ & $13.9 \pm 1.0$\\
 \hline
$\text{M}_{\text{AFM}}$ (pH)\footnote{Maximum available mutual inductance achievable between pairs of flux qubit bodies.} & $1.554$ & $0.582$  \\
 \hline
 $L_q$ (pH)\footnote{Qubit inductance.} & $382.180$ & $142.920$  \\
 \hline
$C_q$ (fF) \footnote{Qubit capacitance.} &  $118.638$ &  $169.388$\\
 \hline
$I_c$ ($\mu A$)\footnote{Qubit critical current.} & $1.994$ & $4.083$\\
 \hline
 Average single qubit thermal width (Ising units) & $0.221$  & $0.117$\\
 \hline
 FM problem freezeout (scaled time) & $0.073$ & $0.015$ \\
 \hline
 Single qubit freezeout (scaled time) & $0.616$ & $0.619$ \\
 \hline
$\Phi^i_{\text{CCJJ}}$ ($\Phi_0$)\footnote{Initial value of the external flux applied to qubit compound Josephson-junction structures at the start of an anneal ($s=0$).}& $-0.624$ & $-0.686$\\
 \hline
$\Phi^f_{\text{CCJJ}}$ ($\Phi_0$) \footnote{Final value at the end of an anneal ($s=1$).} & $-0.723$ & $-0.766$\\
 \hline
Readout time range ($\mu$s)\footnote{Typical readout times for reading between one qubit and the full QPU.} & $18.0$ to $173.0$  & $15.0$ to $48.0$\\
 \hline
 Programming Time\footnote{Typical for problems run on this QPU. Actual problem programming times may vary slightly depending on the nature of the problem.} ($\mu$s) & $\sim 14200$ & $\sim 5500$  \\
 \hline
 QPU delay time per sample ($\mu$s) & $20.5$ & $21.0$  \\
 \hline
 Readout Error Rate \footnote{Error rate when reading the full system.} & $\leq 0.001$  & $\leq 0.001$ \\
 \hline
\end{tabular}
\label{machine_prop}
\end{table}

\end{document}